\documentclass[12pt]{iopart}

\usepackage{epsfig}
\usepackage{graphicx} 
\usepackage{cite}

\newcommand {\bfa} {{\bf a}}
\newcommand {\bfe} {{\bf e}}
\newcommand {\bfk} {{\bf k}}
\newcommand {\bfp} {{\bf p}}
\newcommand {\bfq} {{\bf q}}
\newcommand {\bfr} {{\bf r}}
\newcommand {\bfQ} {{\bf Q}}
\newcommand {\bfD} {{\bf D}}
\newcommand {\bfR} {{\bf R}}

\newcommand {\E} {{\varepsilon}}
\newcommand {\om} {{\omega}}
\newcommand {\Om} {{\Omega}}

\newcommand {\tot} {{\rm tot}}
\newcommand {\pol} {{\rm pol}}
\newcommand {\ord} {{\rm ord}}
\newcommand {\rint} {{\rm int}}

\newcommand {\ee} {{\rm e}}
\renewcommand {\i} {{\rm i}}
\renewcommand {\d} {{\rm d}}
\newcommand {\ap} {{a^{\prime}}}

\newcommand {\out} {{\rm out}}
\newcommand {\rin} {{\rm in}}
\newcommand {\eff} {{\rm eff}}

\newcommand {\talpha} {\tilde{\alpha}}
\newcommand {\np} {n^{\prime}}
\newcommand {\lp} {l^{\prime}}

\begin{document}
\jl{2}

\title{Polarizational bremsstrahlung in non-relativistic collisions}

%\date{  }   % removes data option
%\maketitle  % makes title

\author{A.~V.~Korol$^{\rm a}$, A.~V.~Solov'yov$^{\rm b}$
\footnote{On leave from: Ioffe Physical-Technical Institute, 
Russian Academy of Sciences, St. Petersburg 194021, Russia}
}

\address{
     $^{\rm a}$\, 
     Department of Physics, Russian Maritime Technical University, 
     Leninskii prospect 101, St. Petersburg  198262, Russia} 
  
\address{$^{\rm b}$\, 
     Frankfurt Institute for Advanced Studies,
     Johann Wolfgang Goethe-Universit\"at, 
     60054 Frankfurt am Main, Germany}

%%%%%%%%%%%%%%%% Abstract
\begin{abstract}
We review the developments made during the last decade
in the theory of polarization bremsstrahlung in the non-relativistic domain.
A literature survey covering the latest history of the phenomenon
is given.
The main features which distinguish the polarization bremsstrahlung
from other mechanisms of radiation are discussed and illustrated
by the results of numerical calculations.
\end{abstract}

%%%%%%%%%%%%%%%%%%%%%%%%%%%%%%%%%%%%%%%%%%%%%%%%%%%%%%%%%%
\section{Introduction.\label{Introduction}}

In this paper we review recent developments 
in the theory of polarizational bremsstrahlung (BrS).

\begin{figure}
\begin{center}
\includegraphics[width=12cm,height=4cm,angle=0]{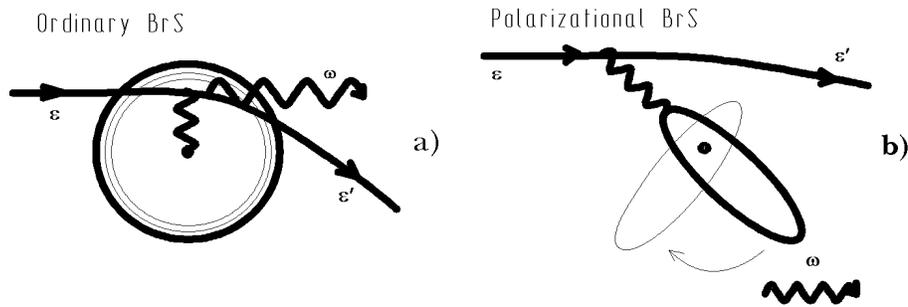}
\end{center}
\caption{
Schematic representation \cite{KorolSolovyov1997} 
of the ordinary and polarizational BrS processes.  
Ordinary BrS is the photon emission of a charged projectile 
accelerated in the static field of the target.  
Polarizational bremsstrahlung mechanism considers the photon
emission of the target electrons, virtually excited by the projectile.  
Virtual excitation of the electrons is equivalent to polarization of the
target.}
\label{BrSPicture.fig}
\end{figure}

Consider the two mechanisms of the photon radiation during a collision 
shown in figure \ref{BrSPicture.fig}. 
On the left-hand side of the figure, ordinary BrS (OBrS) is illustrated.
In this case the emission  of a photon occurs by a charged projectile 
accelerated in the static field of a target. 
This is a well known quantum mechanical process  the basic description
of which can be found in textbooks (see, e.g.~\cite{Akhiezer}).
The right-hand picture in the figure illustrates the
polarizational BrS (PBrS)~\cite{redbook}.
Here, the photon emission  occurs due to the virtual excitations
(polarization) of the target electrons under the action of the field
created by a charged projectile.

%%%%%%%%%%%%%%

The importance and the fundamental character of the OBrS
process has been recognized long ago
(for a review of historical background see~\cite{Pratt1981}).
Since then it has been intensively studied theoretically,
numerically and experimentally in a wide range of the projectile
and the emitted photon energies, different geometries of the emission,
and a variety of atomic and ionic targets.
The  reviews of the results obtained in this field 
include  
\cite{KochMotz,Pratt1984,PrattFeng1985,Nakel1994,Pratt1995}.
The spectral \cite{AtData1} and spectral-angular distributions \cite{AtData2}
of OBrS are tabulated over wide ranges of energies of
projectile electron and emitted photon and for a number 
of atomic targets.

%%%%%%%%%%%%%%%%%%% Two periods: before -after.

The polarizational mechanism of the radiation was recognized relatively 
recently
\cite{BuimistrovTrakhtenberg1975,Amusia1976,PindzolaKelly1976,Amusia1977,
BuimistrovTrakhtenberg1977,Zon1977,WendinNuroh,Zon1979,GolovinskiZon1980} 
where  the first qualitative and quantitative estimates were made 
of the role of the target polarization in forming the BrS spectrum
in an electron-atom collision in the range of photon energies close 
to the atomic ionization potentials.

In the subsequent decade the theory of the effect was developed
further and a number of new phenomena was recognized and described
both analytically and numerically. 
% Electron-atom scattering
An important idea of a close relationship between wide and powerful
maxima in experimentally measured emission spectra of electrons 
\cite{ZimkinaShulakovBrajko1981}
and the giant dipole resonances in photoionization of many-electron
atoms was formulated \cite{AmusiaZimkinaKuchiev1982}.
The effect of a dynamic descreening (`stripping') of a many-electron 
subshell for the photon energy larger than its ionization potential
was formulated \cite{AmusiaAvdoninaChernyshevaKuchiev1985}
and on its basis the additional asymmetry of the experimentally measured 
emission spectrum 
\cite{VerkhovtsevaGnatchenkoPogrebnyak1983} was explained. 
The numerical calculations of the PBrS spectrum and angular distribution  
in collisions of electrons with many-electron atoms were performed
for fast \cite{AmusiaAvdoninaKuchievChernysheva1986,
AvdoninaAmusiaKuchievChernysheva1986} and intermediate energy
\cite{AmusiaKorolChernysheva1990,AmusiaKorol1991} projectiles.
A formalism and specific features of the dipole-photon polarizational
BrS in relativistic collisions of structureless charged particles with 
atoms were reported 
\cite{AmusiaKorolKuchievSolovyov1985,
AstapenkoBuimistrovKrotovMikhailovTrahtenberg1985}.

The theory of PBrS in collisions of charged particles, other than electrons, 
with atoms/ions  was developed. 
In \cite{BuimistrovKrotovTrakhtenberg1980} analytical treatment of the
exactly solvable problem of the PBrS arising in a positron and a proton
collision with a hydrogen atom was given.
It was shown that over wide range of the photon energies
the BrS intensities for both projectiles are of the same order of magnitude.
The PBrS formalism for a bare-ion---atom collision 
\cite{IshiiMorita1984,IshiiMorita1985}
and a bare-ion---hydrogen-ion collision \cite{GonzalezMiragliaGaribotti1988} 
was developed 
and the performed calculations allowed to explain the earlier measured
experimental data on the BrS spectrum of protons.
A more general treatment of the BrS arisen in the collision of two
complex colliders (atoms, ions), which accounts also for the recoil 
effect of the nuclei, was carried out in 
\cite{AmusiaKuchievSolovyov1984,AmusiaKuchievSolovyov1985}.

%% A + A
The important feature of the PBrS mechanism is that it leads to the 
emission in collisions of two electrically neutral objects 
possessing an internal structure.
In \cite{AmusiaKuchievSolovyov1984,AmusiaKuchievSolovyov1985} 
for the first time
it was demonstrated that the intensive BrS emission can appear in 
atom--atom collision. 
In this case the radiation is due to the mutual virtual polarization
of the colliders. 
The total induced dipole moment of the system alters during the 
collision, and
this results in the photon emission.
The general formalism developed in the cited papers 
allowed to express the BrS amplitude via the generalized atomic 
polarizabilities of the colliders.
It was demonstrated that no dipole photon emission appear in symmetric
collisions.
The specific features (due to the Coulomb repulsion) of the PBrS formed 
in ion--ion collision were studied in \cite{AmusiaSolovyov1990b}.
The numerical calculations carried out in \cite{AmusiaKuchievSolovyov1984}
have shown that the radiation intensity in the collision of two neutral
(but different) atoms is comparable to that formed in the collision 
where one of the colliders is substituted with a charged heavy particle 
or an electron of the same velocity.
Later the formalism,  initially developed for non-relativistic atomic
collisions, was extended to the relativistic domain 
\cite{AmusiaKuchievSolovyov1987,AmusiaSolovyov1988,AmusiaSolovyov1990a}.

For several cases which allow for the exact analytical treatment of the PBrS
process the corresponding formulae were derived.
These include the PBrS formed in fast electron--hydrogen atom
collision \cite{DuboiusMaquetJetzke,DuboiusMaquet1989},
in electron/positron collisions with muonic hydrogen 
\cite{AmusiaSolovyov1985,AmusiaGribakinKharchenkoKorolSolovyov1987},
and in the collision of a  charged particle with positronium
\cite{AmusiaKorolSolovyov1986b}.
%% neutron and neutrino
A universal character of the PBrS mechanism, i.e. its (qualitative)
independence on the type of the interaction between a projectile and
a target was demonstrated in 
\cite{AmusiaBaltenkovZhalovKorolSolovyov1986,AmusiaBaltenkovKorolSolovyov1987}
for the processes of neutron-- and neutrino--atom scattering
(see also the eralier publications \cite{Varfolomeev1978,Varfolomeev1980}).
The PBrS manifests itself in nuclear collisions 
\cite{AmusiaSolovyovNuclear1987}, and in nuclear reactions such as
$\alpha$, $\beta$, $\gamma$ decays or nuclear fission process
\cite{Solovyov1989}.

In the papers \cite{AmusiaSolovyov1990a,
AmusiaKuchievSolovyov1985c,AmusiaKorolSolovyov1986a,
AstapenkoBuimistrovKrotov1987}
a theoretical study of the BrS process accompanied by the excitation
or ionization of the target (an `inelastic' BrS) was carried out.
The photon energy intervals were established where the contribution
of the inelastic BrS is small compared to the BrS process without the
change in the target's state.

% Early reviews on PBrS reviews
Some of the results, obtained within the period until the early 90th,
were reviewed in  
\cite{redbook,Amusia1988,Amusia1990}.

During the last decade a systematic quantitative investigation of 
the PBrS has been carried out.  
With the increase of the computer power it has become possible 
to effectively compute the characteristics of BrS for various projectiles,
complex targets (isolated many-electron atoms, atoms in an environment, 
clusters) and over broad ranges of photon energies.
The accurate theoretical predictions on the magnitude of the BrS
cross sections, obtained over ranges of energies and targets 
accessible to experiment 
\cite{Verweyen1996,Quarles1998,QuarlesPortillo1999,PortilloQuarles_PRL},
become available.
The combination of theoretical and numerical tools has allowed 
not only to analyze and test different schemes used to describe 
the scattering process and the dynamic response of the target (these are
two key elements of the PBrS process) but to develop alternative methods as
well.

A general approach to consider the BrS process (both
the ordinary and the polarizational) in collisions with many-electron 
targets is based on a consistent application of quantum mechanics 
and quantum many-body theory. 
The main results obtained by using this approach include:
(a) application of the many-body theory for accurate calculations
of the dynamic generalized polarizability of the target 
\cite{KorolLyalinSolovyov1995a,KorolLyalinSolovyov1996,
KorolLyalinSolovyov1996a,KorolLyalinSolovyovShulakov1996b,
GerchikovKorolLyalinSolovyov1997}
(b) development of the methods for approximate treatment of 
of the dynamic response of the target 
\cite{Korol1992,KorolLyalinSolovyovShulakov1995,
KorolLyalinSolovyovShulakov1996e,KorolLyalinSolovyovShulakov1996c,
KorolObolenskySolovyov1998b,KorolEtAl1999},
(c) calculation of the total BrS spectra of in collisions of 
non-relativistic electrons on many-electron atoms 
over wide range of emitted photon energies including the regions 
of giant resonances 
\cite{KorolLyalinSolovyov1997,KorolLyalinObolenskySolovyov1998,
KorolLyalinSolovyov1999,KorolLyalinSolovyovAvdoninaPratt2002},
(d) theoretical description of the PBrS in collisions of slow atomic
particles and in collisions involving atoms in the excited states
\cite{Solovyov1992,KorolKuchievSolovyov1992,KorolObolenskySolovyov1999},
(e) numerical calculations of inelastic BrS 
\cite{Korol1994c,KorolLyalinObolenskySolovyov2000},
(f) theoretical and numerical description of the BrS process in 
electron-cluster collisions 
\cite{AmusiaKorol1994,ConneradeSolovyov1996,GerchikovIpatovSolovyov1996,
GerchikovSolovyov1997,GerchikovIpatovSolovyov1998},
 (g) the full relativistic description of the BrS in a charged 
particle--atom collision 
\cite{OurRelativisticJPB,OurRelativisticJETP,
KorolObolenskySolovyovSolovjev2002}.

Some of the results presented in the papers cited above
were reviewed by us some time ago \cite{KorolSolovyov1997}.
In full the results obtained by means of the quantum-mechanical
description during the whole period of the PBrS history were
summarized in the book \cite{OurBook2004} (in Russian).
Thus, the motivation for writing the current review is to 
present the results obtained in our group during the latest period 
for a wider community.
In what follows we focus on the achievements made 
in the development of the theory of PBrS formed in 
non-relativistic collisions of various particles with 
isolated many-electron atoms. 
We also included the part devoted to the BrS formed in atom--atom
collisions. 
The reason for this is that although these results were
obtained nearly twenty years ago 
\cite{AmusiaKuchievSolovyov1984,AmusiaKuchievSolovyov1985}
they have not been reviewed in international journals.
The specific features of the polarizational BrS process in electron-cluster 
collisions and those due to the relativistic effects in electron-atom
collisions are not included in this paper but are 
discussed in other reviews in this issue 
\cite{RPC_Clusters2004,RelBrS_2004}.
 
%%%%%%%%%%%%%%%%%%%%%%%%%%%%%%%%%%%%%%%%%%%%%%%%%%%%%%%%%%%%%%%%%
During the last decade another approach to the BrS problem 
(with both channels included) has been developed for theoretical and
numerical description of the process in collision with isolated atoms
and in plasma
\cite{AstapenkoKukushkin1997,Astapenko1999,AstapenkoBureevaLisitsa2000ab,
AstapenkoBureevaLisitsa2000b,AstapenkoBureevaLisitsa2000,
AstapenkoBureevaLisitsa2000c,Astapenko2001}.
It is based on a semi-classical description of the scattering process
and the ordinary BrS channel \cite{KoganKukushkinLisitsa1992}, whereas
the dynamic atomic response is calculated within the 
local density approximation. 
These results were reviewed recently 
\cite{AstapenkoBureevaLisitsa2002,AstapenkoBook2003}, and
are not discussed in detail in the present paper.

Other issues which are left beyond the scope of this review
include the influence of the density effect on the PBrS for energetic 
projectiles penetrating through dense medium 
\cite{BlazhevichEtal1996,Nasonov1998,BlazhevichEtal1999,
KamyshanchenkoNasonovPokhil2001,AstapenkoBuimistrovKrotovNasonov2004}
and
the multiphoton PBrS  \cite{AstapenkoKukushkin1997,Golovinski1988,
AmusiaKorol1993,
KrackeAlberBriggsMaquet1993,KrackeBriggsDuboisMaquetVeniard1994}.

The atomic system of units is used.

%%%%%%%%%%%%%%%%%%%%%%%%%%%%%%%%%%%%%%%%%%%%%%%%%%%%%%%%%%%%%%%%%%%%%
\section{Main features of the polarizational bremsstrahlung process.
\label{MainFeatures}}

In this section we review the peculiar features of 
polarizational BrS which distinguish this mechanism from ordinary BrS
and which manifest themselves in the spectrum 
of the total BrS. 
Focusing on the physical nature of these phenomena and to avoid 
unnecessary complexities, we consider the BrS process of a 
non-relativistic charged structureless particle on a many-electron target
(called an atom, for brevity) within the framework of 
the plane-wave first Born approximation for the 
scattering process and the dipole-photon scheme for the 
description of the photon-atom and photon-projectile interactions.
Such treatment, although quite often being insufficient for
the quantitative description of the process,
allows one to carry out a simple qualitative analysis
and to explain all specific features of PBrS.

Let $\bfp_1$ and $\bfp_2$ denote the momenta of initial and final states 
of the projectile with mass $m$ and charge $e$.
To simplify the formalism we consider the BrS process in the collision with 
a spherically symmetric neutral atom.
This restriction is not too rigid since the effects described below 
also occur in the BrS process on a target with a ground state of  
arbitrary symmetry.

Considering the two mechanisms of the photon emission one derives the
following expression for the total amplitude of the process:
\begin{equation}
f_{\tot} 
= f_{\ord}
+
f_{\pol}
= {4\pi (\bfe\bfq) \over q^2}
\left[ 
{e^2 \over m}\,{Z-F(q) \over \om} 
+ 
e\,\om\, \alpha(\om,q) 
\right],
\label{1.1}
\end{equation}
where $\bfe$ and $\om$ are the photon polarization vector and
energy, $\bfq = \bfp_1 - \bfp_2$ is the momentum transfer.

The first term in (\ref {1.1}) describes the ordinary part of the 
total amplitude.
It is  proportional to $Z-F(q)$, where $F(q)$ 
is the form-factor of atomic electrons and $Z$ is the charge of the nucleus.
Hence,  $f_{\ord}$ is dependent on the static distribution 
of the charge in the atom.
The polarizational amplitude $f_{\pol}$ is expressed via a
generalized atomic dynamic  polarizability $\alpha(\om,q)$ which 
appears as a result of the action of two field on the atom: 
the field of the projectile and the electromagnetic field of the 
emitted photon.

The first feature which distinguishes between the two mechanisms
follows immediately from (\ref {1.1}).  
The amplitude of ordinary BrS is inversely proportional to the mass 
of projectile, while the polarizational part is independent of it. 
The explanation for this fact follows from the basic principles 
of electrodynamics (e.g. ~\cite{Akhiezer}).
During the process of ordinary BrS, it is the projectile that emits the 
photon. 
The intensity of this radiation is proportional to the square of the 
projectile acceleration in the external field of a target. 
In turn, the acceleration is proportional to $1/m$ and this dependence 
manifests itself in $f_{\ord}$. 
In contrast, during the polarizational BrS the projectile serves 
as a source of the external field acting on the atomic electrons, 
and thus the amplitude of this process is almost insensitive 
to the variations of $m$ \cite{redbook}.
Moreover, the intensity of PBrS for a heavy projectile is comparable
and can be even higher than that of an electron of the same velocity
\cite{AmusiaKuchievSolovyov1984}.

Other qualitative differences between the two mechanisms of 
the photon emission one can trace by comparing the dependencies 
of $f_{\ord}$ and $f_{\pol}$ on the photon energy and on 
the momentum transfer.

The OBrS amplitude is a smooth  function of $\om$.
The only peculiarity appears in the soft-photon region $\om\to 0$ 
where a simple perturbative approach, giving an infinite magnitude 
of $f_{\ord}$,  fails to describe the process. 
This phenomenon, known as the "infrared catastrophe", 
had been recognized and understood long ago \cite{Akhiezer}.
The $q$ dependence of $f_{\ord}$ is concentrated in the factor $Z-F(q)$.
The atomic form-factor $F(q)$ is the Fourier image of the 
electron charge distribution and is a monotonically decreasing 
function of $q$.
Qualitatively, the value $F(q)$  defines the number of atomic
electrons inside the sphere of a radius $r\sim 1/q$.
Hence, this function reaches its maximum value at $q=0$ where $F(0) = Z$
and decreases monotonically with the increase of $q$. 
In the case of large $q$,  $\lim_{q \to \infty}F(q)= 0$.
The natural scale to measure the magnitude of $q$ is the inverse radius 
of the target, $R_{\rm at}^{-1}$. 
Thus, the amplitude of OBrS is large for $q>R_{\rm at}^{-1}$ 
while in the region $q\ll R_{\rm at}^{-1}$ it becomes negligibly small. 
Such behaviour has a clear explanation \cite{Pratt1984}.
To radiate a photon via the ordinary mechanism a projectile must penetrate
inside the atom, at a distance $r< R_{\rm at}$, where a strong nuclear
potential $-Z/r$ is less screened by the electron cloud.  
In the opposite limit, when $r\gg R_{\rm at}$, the nucleus
is fully screened by the electrons (in the case of a neutral target)
and the probability for a projectile to get the acceleration and
to radiate vanishes.

The PBrS appears as a result of the alteration of the atomic dipole 
moment induced during the collision. 
There are two external fields - the field the photon
and the Coulomb field of the projectile - which act on the 
atom in this process. 
The dynamic response of the target depends, therefore, on the 
parameters of both fields. 
Formally, it is reflected in the dependence of the generalized dynamic 
polarizability $\alpha(\om,q)$ on two variables.
We use the term 'generalized' when addressing to $\alpha(\om,q)$ in order
to stress the dependence on $q$, and, thus,
to distinguish this quantity from the dipole dynamic polarizability, 
$\alpha_{\rm d}(\om)$, to  which $\alpha(\om,q)$ reduces in the limit of 
small transferred momenta:
\begin{equation}
\lim_{q\to 0}\alpha(\om,q)  = \alpha_{\rm d}(\om)\, .
\label{1.2}
\end{equation}

The dependence on $q$ appears because of the action
of the external Coulomb field of the projectile. 
This field distorts the electrons' orbits and induces a dipole moment 
of the atomic system. 
The dipole polarization of the electron cloud is most pronounced 
if the Coulomb field of the projectile is uniform on the scale of 
$R_{\rm at}$, 
i.e. when the projectile is outside the target, $r\gg R_{\rm at}$. 
These distances correspond to small values of the transferred momentum
$q\ll R_{\rm at}^{-1}$ where, in accordance with (\ref{1.2}),
the PBrS amplitude, as well as the cross section,
can be expressed through $\alpha_{\rm d}(\om)$.
For small distances, $r\ll R_{\rm at}$, the field of the projectile
is almost spherically symmetric.
Therefore, it induces a small dipole moment on the target. 
Hence, in contrast to the ordinary BrS process, it is the 
large distances between the projectile and the target which 
are of the most importance for the PBrS mechanism
\cite{Zon1977,AmusiaZimkinaKuchiev1982}.

%Therefore, although in the case of light projectiles (electron, positron) 
%two mechanisms of BrS formation must be treated simultaneously,
%it is possible, in principle,  to distinguish the
%photons emitted via the polarizational mechanism from those which are 
%formed through the ordinary one.  
%To do this it is necessary to detect the emitted photon in coincidence  
%with the scattered particle. 
%Then the "polarizational" photons will be observed together with the 
%projectile scattered at small angles (corresponding to the transferred 
%momenta $q<R_{\rm at}^{-1})$, while large scattering angles correspond 
%to the emission of the "ordinary" photon. 
%This consideration is valid for neutral atoms but does not hold for ions, 
%since in the latter case a long-range Coulomb potential leads to the 
%emission of the photon via the ordinary mechanism for small-angle 
%scattering of the projectile.

The $\om$ dependence of  $\alpha(\om,q)$ reflects  the ability of 
the electron cloud to be dynamically polarized by an external
electromagnetic field of a given frequency. 
In a many-electron atom the electrons are distributed among 
the atomic subshells. 
Each subshell is characterized by an ionization potential $I$.  
In terms of classical mechanics  this corresponds to the
frequency of the rotation of electrons  of a given subshell around the 
nucleus.
Using this analogy one may say that the dynamic response of the electron 
cloud to the external field increases for those $\om$ which are 
close to the ionization thresholds of the target subshells. 
Therefore, in the region
\begin{equation}
I_1 < \om < I_{\rm 1s},
\label{1.3}
\end{equation}
where $I_1$ and  $I_{\rm 1s}$ stand, respectively,
for the ionization potentials of the outermost shell and of the $1s$ shell,
the function $\alpha(\om,q)$ is non-monotonic with extrema 
in vicinities of the ionization potentials.

This happens, in particular, when the photon energy lies within the 
region of a giant dipole resonance of the photoionization cross section 
of a many-electron atomic subshell \cite{AmusiaPhoto}.
For the first time wide maxima in the emission spectra  
were observed  experimentally in electron scattering from (solid) 
Ba, La and Ce \cite{LiefeldBurrChamberlain1974,ChamberlainBurrLiefeld1974}.
Later these were explained theoretically
\cite{WendinNuroh} by relating the maxima to 
the virtual excitations of the $3d$-subshell electrons.
In \cite{ZimkinaShulakovBrajko1981} a powerful maximum was observed
in the emission spectrum in electron-La collision.
In the subsequent paper \cite{AmusiaZimkinaKuchiev1982}
for the first time an important conclusion was drawn about
the common nature of the giant resonances in photoionization and
those in PBrS spectra.
To reveal this similarity one recalls that at large distances
between the projectile and the atom the amplitude of the PBrS process  
is proportional to the the dipole dynamic polarizability, 
$\alpha_{\rm d}(\om)$. 
The imaginary part of this quantity is related to the photoionization
cross section  $\sigma_{\gamma}(\om)$  through (e.g. \cite{Land4}):
\begin{equation}
{\rm Im}\,\alpha(\om) = {c \over 4\pi\om}\,
\sigma_{\gamma}(\om),
\label{11}
\end{equation}
where $c\approx 137$ is the velocity of light.
Since the OBrS amplitude is real (see (\ref{1.1})), 
the modulus square of the imaginary part  of $f_{\pol}$ 
enters the total BrS cross section as an additive term.
Therefore, a maximum in $\sigma_{\gamma}(\om)$ manifests itself 
in the BrS spectrum as well reflecting the collective nature 
of the dynamic response of atomic electrons.

Although based on the assumption that main contribution to 
$f_{\pol}$ comes from the region of large distances $r\gg R_{\rm at}$, 
which does not always lead to a correct quantitative result, 
the qualitative arguments of \cite{AmusiaZimkinaKuchiev1982} 
provided a clear physical explanation of the nature of powerful
maxima in emission spectra.
The experiments, carried out later, supported the theoretical 
prediction.
The maxima in the BrS spectra were measured for Ba and several
rare-earth elements \cite{ShulakovZimkinaBrajkoEtAl1983,
ZimkinaShulakovBrajkoEtAll1984},
for La and for atoms from the lanthanum group 
\cite{ZimkinaShulakovBrajkoEtAl1984a},
for Xe \cite{VerkhovtsevaGnatchenkoPogrebnyak1983,
VerkhovtsevaGnatchenkoZonNekipelovTkachenko1990}
and for Ba \cite{Verweyen1996}.
In all these experiments, performed for various energies
of the incoming electron (ranging from several hundreds of eV up
to several keV), the powerful BrS maxima were observed for photon
energies within the ranges of the giant resonances in the 
photoionization cross section of the $4d$ subshells.

The existence of the maxima in the BrS spectrum due to 
a strong effect of polarization of many-electron atomic subshells
was confirmed by a number of independent theoretical calculations
\cite{AmusiaAvdoninaChernyshevaKuchiev1985,
AmusiaAvdoninaKuchievChernysheva1986,AvdoninaAmusiaKuchievChernysheva1986,
AmusiaKorolChernysheva1990,AmusiaKorol1991,AmusiaKuchievSolovyov1984,
KorolLyalinSolovyov1995a,KorolLyalinSolovyov1996,
KorolLyalinSolovyovShulakov1995,
KorolLyalinSolovyovShulakov1996b,KorolLyalinSolovyovShulakov1996c} 
which were carried out within the frameworks of various models.

In general, in the whole $\om$-region defined by (\ref{1.3}),
a highly non-monotonic behaviour of the generalized polarizability of 
a many-electron atom results in a series of peculiarities 
(maxima, minima, cusps) in the total BrS spectrum.
The important role of the polarizational mechanism in 
forming the total BrS spectra 
over a wide range of photon energies was analyzed in 
\cite{KorolLyalinSolovyov1997,
KorolLyalinObolenskySolovyov1998,
KorolLyalinSolovyov1999,
KorolObolenskySolovyov1999,Romanikhin2002}.

For the photon energies noticeably higher than the $1s$ 
ionization potential, $\om \gg I_{\rm 1s}$, the $\om$ dependence 
of $\alpha(\om,q)$ is much like that for the cloud of free electrons. 
The leading term in the expansion of $\alpha(\om,q)$ in powers of
$\om/I_{1s}$ reads: 
\begin{equation}
\alpha(\om,q)\approx -{F(q) \over \om^2}\, .
\label{1.4}
\end{equation}
As a result, for a fast electron ($m=1$, $e=-1$) 
the total amplitude reduces to the BrS amplitude 
on a bare nucleus \cite{BuimistrovTrakhtenberg1977}:
\begin{equation}
f_\tot 
\approx 
{4\pi (\bfe\bfq) \over q^2}
\, {Z \over \omega}\, .
\label{1.5}
\end{equation}
This shows that for large $\om$
the atomic electrons do not participate in the screening of the nucleus and 
do not contribute to the BrS cross section.
Thus, in the region $\om \gg I_{\rm 1s}$ the polarizational channel 
results in a (dynamic) de-screening of the nucleus.

The physical reason for this effect 
(following \cite{AmusiaAvdoninaChernyshevaKuchiev1985} we use
the term `stripping' effect) 
is that, for $\om \gg I_{\rm 1s}$, 
the electrons of all atomic subshells may be treated as free ones
\cite{BuimistrovTrakhtenberg1977}. 
If the incident electron is also free (the Born approximation) then 
there is no dipole radiation by a system of free electrons.

These arguments allow one to construct an approximate expression
for the total BrS amplitude for photon energies lower than the 
$1s$-shell ionization threshold \cite{AmusiaAvdoninaChernyshevaKuchiev1985}.
For a fixed value of $\om$ the target electrons can be divided into 
two groups, the 'inner' and the 'outer' electrons.
The former are those whose binding energies, $I_{\rin}$, exceed $\om$, 
and, therefore, their orbits are not distorted noticeably
by an external electromagnetic field of a frequency $\om$.
Hence, the inner electrons do not contribute to the amplitude of the PBrS.
The outer electrons have the binding energies  
$I_{\out}$ less than $\om$.
Under the action of the field they behave as free electrons,
and their contribution to $f_{\pol}$ can be described by (\ref{1.4}) 
where $F(q)$ must be substituted with the form-factor of the outer 
electrons, $F_{\out}(q)$. 
As a result the total BrS amplitude acquires the form
\begin{equation}
f_{\tot} 
\approx 
{4\pi (\bfe\bfq) \over q^2}
\, 
 {Z-F_{\rin}(q) \over \om}\, ,
\label{1.1a}
\end{equation}
where $F_{\rin}(q)$ stands for the form-factor of the inner electrons.
This expression demonstrates that the outer electrons do not participate 
in the screening of the nucleus
(or, in other words, the nucleus is 'stripped' by a total
 number $N_\out$ of the outer electrons).
The physical reason for this partial 'stripping' is as formulated above:
for $\om \gg I_{\out}$ the outer electrons
can be considered as free and, therefore, there
is no dipole-photon emission by the
system 'projectile electron + the outer electrons'.

Although the 'stripping' approximation does not account for the 
specific details of the PBrS amplitude near each subshell threshold,
it allows one to estimate the behaviour
of the smooth background of the BrS cross section curve.
In particular, on  its basis  the asymmetry of the giant resonances
in the experimentally measured emission spectra 
\cite{ShulakovZimkinaBrajkoEtAl1983,
ZimkinaShulakovBrajkoEtAl1984a,
VerkhovtsevaGnatchenkoPogrebnyak1983} was explained.
Namely, from (\ref{1.1a}) follows that the difference between
two values of the function $\om f_{\tot}$ calculated for 
(a) $\om \gg I_j$ ($I_j$ is the ionization potential of 
the $j$th subshell), and for (b) $\om \ll I_j$,
is proportional to the form-factor $F_j(q)$ of the subshell which can be
estimates as $F_j(q)\approx N_j$, where
$N_j$ is the number of the electrons in the subshell.
Therefore, the difference between the two amplitudes is proportional 
to $N_j$, 
and the increase in the cross section is 
$\sigma(\om\gg I_j) - \sigma(\om\ll I_j) \propto N_j^2$
\cite{AmusiaAvdoninaChernyshevaKuchiev1985}.

This qualitative explanation was confirmed by numerical calculations 
\cite{AmusiaAvdoninaKuchievChernysheva1986} carried out
within the framework of the non-relativistic Born approximation. 
Later the 'stripping' was generalized, going beyond 
the Born approximation \cite{Korol1992,KorolEtAl1999,AvdoninaPratt1999}.

When the photon energy becomes small compared with the first ionization
potential of the target  one can expect the decrease of the
contribution of the polarizational BrS channel into the total spectrum.
From (\ref{1.1}) follows that for $\om\ll I_1$ the PBrS amplitude 
behaves as $f_{\pol} \propto \om \alpha(0,q) \sim \om \alpha_{\rm d}$, 
where $\alpha_{\rm d}$ is a static dipole polarizability of the target.
In contrast, the OBrS term increases.
For the ratio of the two terms 
$f_{\pol}/f_{\ord}\sim m\om^2 \alpha_{\rm d}/e$
vanishes as $\om$ goes to zero.
For low but non-zero values of $\om$ the contribution 
of the polarizational channel strongly depends on the magnitude of the
static polarizability. 
The higher the magnitude of  $\alpha_{\rm d}$ is,  the wider is the
$\om$ interval where the contribution of $f_{\pol}$ might be noticeable.

These arguments are valid also beyond the range of validity
of the first Born approximation, on the basis of which 
(\ref{1.1}) was derived.
In the low-frequency limit the OBrS amplitude is expressed in terms
of the amplitude of elastic scattering, $f_{\ord} \propto f_{\rm el}/\om$
(e.g., \cite{Land4}).
On the other hand, the process of PBrS occurs most effectively
at large distances between the projectile and the target, 
where the wavefunction of the projectile is not
affected strongly by the potential, and one can
use the Born approximation to describe the polarizational channel
even for low-energy projectiles.
Therefore, the ratio of the two amplitudes still 
is proportional to $\om^2 \alpha_{\rm d}$.
In \cite{Zon1979} the role of the polarizational channel
was studied in the inverse BrS process (i.e. BrS absorption)
for slow electrons scattered from atoms.
It was demonstrated that for an $e^{-}$-Ar scattering the polarizational 
mechanism changes noticeably the absorption coefficients, whereas
for an $e^{-}$-Ne scattering its influence is much less.
This quantitative effect is due to large difference in the  
static polarizabilities: 
$\alpha_{\rm d}^{\rm Ar} = 11.10$ a.u. and 
$\alpha_{\rm d}^{\rm Ar} = 2.66$ a.u.
\cite{RatzigSmirnov}.
Rather strong effect of the target polarization on the BrS
spectra was reported recently \cite{HammerFrommhold2001} for low-energy
($\E_1=0.4\dots 3.5$ eV) electron--rare-gas atom collisions.

%%%%%%%%%%%% Molecular orbital X-rays and excited hydrogen 

In general, in the collisions of slow heavy particles with atoms 
both the ordinary and the polarizational mechanisms fail
to describe adequately the  radiation spectrum.
In such processes another mechanism, known as molecular 
orbital radiation, becomes important (see, e.g., \cite{GreinerReinhardt}).
Nevertheless, the polarizational BrS is important in asymmetric slow 
collisions of atoms and ions in the region  of large impact parameters 
\cite{Solovyov1992}.
The intensity of the OBrS is negligibly small due  
the large masses of the colliders.

However, the situation with the radiation formed in the slow charged
particle--excited hydrogen collision is somewhat different, compared 
with PBrS and aslo with molecular orbital radiation.
More accurately, in addition to these types of radiation there is a 
peculiar source of low-frequency photon emission 
\cite{KorolKuchievSolovyov1992}. 
In this case the radiation is generated by the rotating dipole moment of 
the hydrogen during the collision.
The specific feature of the hydrogen atom introduces the 
linear Stark effect (see, e.g., \cite{Landau3}).
The electric field of the projectile splits the initially degenerated
levels of the excited hydrogen. 
Atomic states with a given principal number form a Stark multiplet. 
The components of the multiplet already possess a dipole moment. 
The vector of this dipole moment rotates following the movement of 
the projectile, and the radiation appears as a result of this rotation. 
We stress that this mechanism of BrS is intrinsic for systems with 
a linear Stark effect. 
This is also a distinguishing feature from the `real' 
PBrS which also appears because of the alteration of the target's
dipole moment.
In the latter case it is really the induced dipole moment 
intrinsic for systems with a quadratic Stark effect. 
As a result the character of these spectra at low frequencies are 
quite different \cite{KorolKuchievSolovyov1992}.

%%%%%%%%%%%% Non-relativistic  inelastic BrS for atoms  %%%%%%%%%%%%%%%%%

Once the internal dynamic structure of a target is taken into account,
the next logical step is to consider the radiative
processes which are accompanied by the excitation or ionization of the 
target.
Following \cite{AmusiaKuchievSolovyov1985c} we call BrS processes 
of this type `inelastic' BrS  contrary to the `elastic' one, when 
the target remains in its ground state after the collision. 
Within the framework of the same approximations as above
the amplitude of the scattering process 
accompanied by the real atomic transition from the initial ground state 
$0$ to the final state $m$ may be written as follows:
\begin{eqnarray}
f^{(m)}_{\tot}  
=  f^{(m)}_{\ord}  + f^{(m)}_{\pol} 
=
{4\pi \over q^2}
\left[
-{ {\bfe\bfq} \over \om}\,{e^2 \over m}\,
F_{m0}(\bfq)
+ 
e \om  A_{m}(\bfe,\om,\bfq)
\right].
\label{1.6}
\end{eqnarray}
Here $F_{m0}(\bfq) = \langle m\left|\exp({\rm i}\bfq\bfr) \right|0\rangle$
is a non-diagonal form-factor and the function
$A_{m}({\bfe},\om,{\bfq})$ is defined as
\begin{equation}
A_{m}(\bfe,\om,\bfq) 
=
\sum_n 
\left\{
{
\langle m\left | \bfe\hat{\bfp} \right|n\rangle \, F_{n0}({\bfq})
\over \om_{nm} - \om - \i\,0
}
+
{
F_{mn}(\bfq)\, \langle n\left| \bfe\hat{\bfp} \right|0 \rangle
\over \om_{n0} + \om - \i\,0
}
\right\},
\label{1.8}
\end{equation}
where the sum is carried out over the whole set  of 
virtually excited states of the target, $n$, 
$\om_{nm}$ and $\om_{n0}$ are the transition energies,
$\hat{\bfp}$ is the momentum operator. 
In the case $m=0$,  i.e. the elastic BrS, this  function reduces to 
$(\bfe\bfq)\,\alpha(\om,q)$ in accordance with (\ref{1.1}).

It is important to establish the contribution of inelastic
channels to the total emission spectrum.
This is not purely of theoretical interest since experimentally 
it is quite difficult to separate elastic and inelastic channels. 
To do this it is necessary to observe the final state of 
the target with simultaneous detection of the photon.

It has been demonstrated that over a wide region of the photon frequencies,
the elastic channel dominates over the inelastic one  in the total BrS 
spectrum for both heavy  
\cite{AmusiaKuchievSolovyov1985c,AstapenkoBuimistrovKrotov1987}
and light 
\cite{AmusiaKorolSolovyov1986a,AstapenkoBuimistrovKrotov1987}
projectiles scattered on a many-electron atom.
Semi-quantitatively, the cross sections of the elastic BrS, of both the
ordinary and the polarizational nature, exceed those of the inelastic by 
a factor $Z$.
The explanation is as follows 
\cite{AmusiaKuchievSolovyov1985c,AmusiaKorolSolovyov1986a}.
During the elastic BrS the contributions of each atomic electron to the 
polarizational part of the total amplitude (\ref{1.1}) are coherent, 
as in Rayleigh scattering of light.  
Considering the case of a neutral atom and having in mind that the
ordinary part of the elastic BrS spectrum is approximately proportional
to the nuclear charge squared, one finds that the total elastic cross 
section is proportional to $Z^2$. 
In contrast, during the inelastic BrS the contributions of each 
electrons must be summed in the cross section rather than in the 
amplitude (\ref{1.6}).  
Hence, the inelastic BrS cross section is proportional to $Z$ and is 
parametrically small in the case of a many-electron target, when $Z\gg 1$. 

The region of the photon frequencies, in which the above mentioned
coherence effect plays an essential role, is estimated as
\cite{AmusiaKuchievSolovyov1985c,AmusiaKorolSolovyov1986a,
AstapenkoBuimistrovKrotov1987}
\begin{equation}
I_1 < \om < {v_1 \over R_{\rm at}}
\label{1.9}
\end{equation}
where $v_1$ is the initial velocity of projectile.
Beyond the region of coherence inelastic BrS becomes more important.
An exception of this rule occurs in collisions of fast heavy charged 
particles with atoms/ions collisions  (as well as in atom-atom, ion-atom 
and ion-ion) in the region of high photon frequencies.
In this region the  process of inelastic BrS has a threshold, 
which is equal to $\om_{\max}\approx v_1^2/2$. 
However, the elastic BrS takes place at higher
energies, up to $\om \geq 2v_1^2$ , dominating
in this region in the total photon emission spectrum.
Therefore, the photon energy range  $v_1^2/2 \le \om \le 2v_1^2$
is convenient for the observation of the elastic PBrS.
We note that in this $\om$ region there is a peculiar feature in 
the spectrum of PBrS 
\cite{KorolLyalinObolenskySolovyov2000} similar to that which occurs
in inelastic scattering, where it is known as the Bethe ridge 
\cite{Landau3}. 
In more detail we discuss this phenomenon in section \ref{chapter5.2}.

The numerical comparison of the relative role of elastic and 
inelastic channels in proton-atom collisions 
was performed in 
\cite{KorolLyalinObolenskySolovyov2000,AstapenkoBureevaLisitsa2000}.

In electron/positron--many-electron atom scattering elastic BrS dominates 
parametrically over the inelastic one in the region (\ref{1.9}).
However, if one is interested in accurate data on the total BrS cross 
section it is necessary to include the inelastic channels into the 
computational scheme.
Up to now, mostly due to technical difficulties,
numerical investigations of the role of inelastic BrS have not been
as extensive as in the elastic case.
The achievements in this field include the model theoretical study
carried out in 
\cite{VerkhovtsevaGnatchenkoZonNekipelovTkachenko1990,Zon1995}
in connection with the experimental data 
on the intensity of the BrS spectrum in $e^{-}+$Xe collision
as a function of the incoming electron energy $\E_1$ 
\cite{VerkhovtsevaGnatchenkoZonNekipelovTkachenko1990,
TkachenkoGnatchenkoVerkhovtseva1995} 
(see also 
\cite{GnatchenkoTkachenkoVerkhovtseva2002a,
GnatchenkoTkachenkoVerkhovtseva2002b}
for the experimental data in $e^{-}+$Ar collision).

For low-$Z$ targets the absolute magnitudes of both terms from
(\ref{1.1}), $f^{\ord}$ and $f^{\pol}$, and the terms
$f_{\ord}^{(m)}$ and $f_{\pol}^{(m)}$ from (\ref{1.6}) are all of 
the same order.
In this case, it is the charge of the projectile which introduces 
peculiarities in the total radiative spectrum. 
It can be shown that, both for high frequencies of the photon, 
$\om\gg v_1/R_{\rm at}$ 
\cite{BuimistrovKrotovTrakhtenberg1980,AmusiaKuchievSolovyov1985c}
and for the low ones, $\om\ll 1\ $ 
\cite{AmusiaKuchievSolovyov1985c,Korol1994c} the role of inelastic channels 
is negligibly small compared with the elastic BrS in the 
case of electron scattering, while for the positron-atom collision 
the situation is the opposite. 
It occurs mainly because of the difference in the behaviour of the 
interference between the ordinary and polarizational amplitudes in 
inelastic BrS.
The interference is negative in the case of the electron projectile 
and positive in the positron case.
The simplest way to trace this effect is to consider the inelastic BrS 
amplitude (\ref{1.6}) in the limit of high photon frequencies, as 
was done above for elastic BrS. 
For $\om\gg I_{\rm 1s}$ the polarizational inelastic amplitude 
reduces to
$f^{(m)}_{\pol}\approx-(4\pi\, e/ q^2)\,(\bfe\bfq/ \om)\,F_{m0}({\bfq})$.
Using this result in  (\ref{1.6}) one notices that for a projectile
electron ($e=-1$) the ordinary and polarizational terms
cancel each other out, while for a positron ($e=+1$) the effect 
is opposite.

%%%%%%%%%%%% Methods used                                                   
To conclude this part of the paper we review the theoretical
approaches used to describe the BrS process in non-relativistic
collisions.
These can be subdivided into three parts: (a) methods applied to
analyze the scattering process,
(b) models used to describe the interaction with the photon, and 
(c) the approximations used to describe the dynamic atomic response.

The variety of theoretical approaches used  to describe the 
scattering process range from the first-order plane-wave 
Born approximation to more sophisticated ones. 
Since the OBrS phenomenon has much longer history these methods were
first tested in application to this process and later on were used 
in the PBrS problem.
In application to the BrS problem in electron--atom scattering 
the models beyond the plane-wave Born approximation, used in
both the non-relativistic and the relativistic domains 
include the corrections due to the Elwert factor and
its modifications~\cite{AvdoninaPratt1999,LeeKisselPrattTseng1976,
AvdoninaPratt1995}, the use of Sommerfeld-Maue functions 
\cite{Land4,LeeKisselPrattTseng1976},
the approaches based on the use of classical 
\cite{KimPratt1987,FlorescuObolenskyPratt} and semi-classical 
\cite{AstapenkoBureevaLisitsa2002,KoganKukushkinLisitsa1992} 
theories.
The best available results have been obtained using the 
distorted partial-wave expansion  of the projectile wavefunction.
This scheme has been applied to study the ordinary BrS process
of non-relativistic projectiles in the dipole-photon
approximation \cite{Zhdanov1978,Tseng1989}.
The most adequate description of the process
has been obtained by applying the DPWA and using the multipole expansion 
for the projectile-photon interaction operator 
\cite{ShafferPratt1997,Tseng,TsengPratt1971,KellerDreizle1997}.

In many papers on the PBrS problem the non-relativistic Born 
approximation was used for both light (a positron, an electron) and
heavy (a proton, an ion) projectiles.
Although the range of validity of the Born approximation for PBrS is
larger than for OBrS, to obtain more accurate data on the total BrS
cross sections of a light projectile it is necessary to go beyond this
scheme. 
Therefore, the non-relativistic DPWA formalism was
developed 
\cite{AmusiaKorol1989,AmusiaKorolChernysheva1990,
AmusiaKorol1991,AmusiaKorol1992,KorolLyalinSolovyov1995a}
and applied to calculate
the cross sections ${\d \sigma}$ and ${\d^2 \sigma}$ 
over a broad spectral range 
\cite{KorolLyalinSolovyov1996,KorolLyalinSolovyovShulakov1996b,
KorolLyalinSolovyov1997,KorolLyalinObolenskySolovyov1998,
KorolLyalinSolovyov1999,KorolObolenskySolovyov1999,Romanikhin2002}
for non-relativistic electrons scattered on many-electron atoms.
The partial-wave approach was also used to study the PBrS process
of slow electrons \cite{Zon1979,GolovinskiZon1980,Kurkina1,Kurkina2}.

In most of the papers the dynamic atomic response to the joint 
actions of the field of the projectile and of the radiation field 
was treated within the frame of the non-relativistic dipole-photon 
theory
(with exception for several papers mentioned below).
Even within this framework the accurate calculation of 
the generalized polarizability $\alpha(\om,q)$ is not a simple 
task.
Apart from the case of a hydrogen atom (or hydrogen-like ion)
where the analytical evaluation is possible
\cite{BuimistrovTrakhtenberg1975,DuboiusMaquetJetzke,
KorolObolenskySolovyov1999} 
one has to use more sophisticated approaches to calculate this quantity.
The methods known to us include the the Hartree-Fock based 
calculations with 
\cite{AmusiaAvdoninaChernyshevaKuchiev1985,
AmusiaAvdoninaKuchievChernysheva1986,
AvdoninaAmusiaKuchievChernysheva1986,
KorolLyalinSolovyov1995a,
KorolLyalinSolovyov1996,
KorolLyalinSolovyov1996a,
KorolLyalinSolovyovShulakov1996b,
KorolLyalinSolovyovShulakov1995,
KorolLyalinObolenskySolovyov1998} and without 
\cite{AmusiaKorolChernysheva1990}
the inclusion of many-body corrections,
the approach based on the local-density approximation 
(see \cite{AstapenkoBureevaLisitsa2002} and references therein).
Another semi-empirical approach for effective and quite accurate 
calculation of  $\alpha(\om,q)$ for complex systems
was proposed recently 
\cite{KorolLyalinSolovyovShulakov1995,KorolLyalinSolovyovShulakov1996e,
KorolLyalinSolovyovShulakov1996c}.
This method as well as the approach  \cite{KorolObolenskySolovyov1998b}
based on the use of the non-relativistic Coulomb Green function
and valid for the calculation of $\alpha(\om,q)$ in the vicinities of
K- and L-shells 
are described in sections \ref{section:results} and \ref{chapter5.2}.

Beyond the dipole-photon approximation the PBrS was considered in
collisions of a non-relativistic heavy projectile with  many-electron
atom 
\cite{AstapenkoBuimistrovKrotovMikhailovTrahtenberg1985,
GonzalezPacherMiraglia1988,retard2,AmusiaKuchievSolovyov1987,
AmusiaSolovyov1988}.
In these papers the corrections of the order ${k R_{\rm at}}\ll 1$ were
considered and it was demonstrated that they lead to
the additional modification of the angular distribution of the 
radiation.
More systematic analysis of the non-dipolar corrections has become available
recently within the framework of the full relativistic description of
the PBrS process \cite{OurRelativisticJPB,OurRelativisticJETP}.

In what follows we discuss in some more the formalism related to the
PBrS problem and present the results of numerical calculations 
of the BrS spectrum  to illustrate the main features of the process
described above.

%%%%%%%%%%%%%%%%%%%%%%%%%%%%%%%%%%%%%%%%%%%%%%%%%%%%%%%%%%%%%%%%%%%%%%
\section{Bremsstrahlung of electrons on atoms.
\label{section:results}}

The differential cross section, which characterizes the spectral
distribution of the radiation, is given by
\begin{eqnarray}
\d \sigma_{\tot}(\om)
&\equiv
\om { \d\sigma_{\tot} \over {\rm d}\om } 
=
{\om^4 \over (2\pi)^4 c^3} {p_2 \over p_1}
\sum_{\lambda} 
\int \d\Om_{\bfp_2}
\int \d\Om\,
| f_{\tot} |^2
\nonumber\\
&=
\d \sigma_{\ord}(\om)
+
\d \sigma_{\pol}(\om)
+
\d \sigma_{\rint}(\om)
\, .
\label{1}
\end{eqnarray}
The integration is carried out over the directions of propagation of
the scattered particle ($\d\Om_{\bfp_2}$) and the emitted photon
($\d\Om$), 
the summation is over the photon polarizations.
In a general case,  the amplitude $f_{\tot}$ includes the ordinary and 
the polarizational parts.
Therefore, there are three terms appearing in the cross section 
of the process.
In (\ref{1}) $\d \sigma_{\ord}(\om) \propto | f_{\ord}|^2$
and $\d \sigma_{\pol}(\om) \propto | f_{\pol}|^2$
stand for the cross sections of the OBrS and PBrS processes, 
and $\d \sigma_{\rint}(\om)$ is the interference term proportional 
to Re\,($f_{\pol}^{*}f_{\ord}$). 
Note that $\d \sigma_{\rint}(\om)$ can be of either sign.
In the case of a heavy projectile the ordinary BrS can be
neglected and, thus, $f_{\tot}\approx f_{\pol}$ and 
$\d \sigma_{\tot}(\om) \approx \d \sigma_{\ord}(\om)$,
For a light projectile it is necessary to retain all the  terms
in the amplitude and the cross section.

In this section we consider the case of electron--atom scattering.
An adequate description of the BrS emission formed in the intermediate
energy electron-atom collision is obtained by using the 
distorted partial-wave approximation (DPWA) 
\cite{AmusiaKorolChernysheva1990,AmusiaKorol1992}.
In the lowest order of the non-relativistic perturbation theory in the
electron--dipole-photon interaction and in the Coulomb interaction
$\hat{V}=\sum_a |\bfr - \bfr_{\rm a}|^{-1}$, between the incident ($\bfr$)
and atomic ($\bfr_{\rm a}$)  electrons (the sum is carried out over 
all atomic electrons), 
the amplitudes $f_{\ord}$ and  $f_{\pol}$ are given by the expressions:
\begin{eqnarray}
\fl
\qquad
f_{\ord} 
& = 
\langle \bfp_2^{(-)} | \bfe\bfr | \bfp_1^{(+)}\rangle
\label{3}\\
\fl
\qquad
f^{\pol} 
& = 
-
\sum_{n} 
\left[
{\langle 0| {\bfe\bfD}| n \rangle 
\langle\bfp_2^{(-)}\, n | \hat{V}| \bfp_1^{(+)}\, 0\rangle
\over \om - \om_{n0}  + \i 0 }
 -
{\langle\bfp_2^{(-)}\, 0 | \hat{V} | \bfp_1^{(+)}\, n\rangle
\langle n| {\bfe\bfD} | 0 \rangle
\over \om + \om_{n0}  }
\right]\,,
\label{4}
\end{eqnarray}
where $|\bfp_1^{(+)}\rangle $ and $|\bfp_2^{(+)}\rangle$  
are the wavefunctions of the
incident and the scattered electrons with asymptotic momenta
${\bfp}_1$  and ${\bfp}_2$, respectively.
The `$\pm$' superscripts correspond to the out- (`+') and 
to the in- (`-') scattering states, the 
DPWA expansions of which is
\begin{equation}
\left|{\bfp}_{j}^{(\pm)}\rangle\right. 
= 
4\pi \sqrt{{\pi \over p_{j}}}
\sum_{l m} \i^l\, 
\exp(\pm \i\delta_{l}(p_j))\,
{P_{\nu_{j}}(r) \over r}\, 
Y^{*}_{l m}({\bfp}_{j}) Y_{l m}({\bfr})\, .
\label{5}
\end{equation}
Here $\delta_l(p)$ are the phaseshifts,
the notation $\nu$ stands for a set of quantum numbers $(p, l)$.  
The radial wave functions $P_{\nu}(r)$ satisfy the 
Schr\"odinger equation with the `frozen'  core.

Vector ${\bfD}$ in (\ref{4}) is the operator of the dipole interaction of
the atomic electrons with the electromagnetic field, 
$\om_{n0} = E_n - E_0$ is the energy of the atom's transition from 
the ground state $0$ to the virtually excited state $n$ 
(including excitations into the continuum).

The PBrS amplitude can be expressed in terms of the generalized dynamic
polarizability $\alpha(\om,q)$.
Omitting the details (see \cite{KorolLyalinSolovyov1995a})
we present the result:
\begin{equation}
f_{\pol} 
= 
{\i \over 2\pi^2}\,
\int \d \bfQ\,
{\bfe\bfQ \over Q^2}\,
\langle \bfp_2^{(-)}\left|\ee^{-\i\bfQ\bfr}\right|\bfp_1^{(+)}\rangle
\, 
\alpha(\om,Q)\,,
\label{12}
\end{equation}
where the integration is carried out over the entire space of the 
vector $\bfQ$.
This form of representation provides a straightforward reduction 
to the Born limit of $f_{\pol}$.
Indeed, substituting the distorted waves 
$|{\bfp}_{1,2}^{(\pm)}\rangle$ with the wavefunctions for a free particle, 
$|\tilde{{\bfp}}_{1,2}^{(\pm)}\rangle = \exp(i \bfp_{1,2}\bfr)$
one derives the expression for
$f_{\pol}$ within the framework of the plane-wave Born approximation
(see the second term on the right-hand side of \ref{1.1}).

Using the DPWA series (\ref{5}) in (\ref{3}), (\ref{12}) and then 
in  (\ref{1}) one derives the partial-wave series for 
the cross section $\d\sigma_{\tot}(\om)$:
\begin{equation}
\d\sigma_{\tot}(\om)
=
{32 \pi^2 \over 3}\,{\om^4 \over c^3 p_1^2}\,
\sum_{l_1 l_2}
l_{\max}\,
\left|R_{l_2l_1}^{\ord}+R_{l_2l_1}^{\pol}\right|^2\, ,
\label{6}
\end{equation}
where $l_2=l_1\pm1$ in accordance with the dipole selection rules,
and $l_{\max}=\max\{l_1,l_2\}$.
The partial amplitudes of ordinary, $R_{l_2 l_1}^{\ord}$,
and polarizational, $R_{l_2 l_1}^{\pol}$, BrS are defined as follows:
\begin{eqnarray}
R_{l_2 l_1}^{\ord} 
&= 
\langle \nu_2\parallel r \parallel\nu_1\rangle
\label{16} \\
R_{l_2 l_1}^{\pol} 
&=
-
{2 \over \pi} \int_0^{\infty}
dQ\, Q\,
\langle \nu_2\parallel j_1(Qr) \parallel\nu_1\rangle
\, \alpha(\om,Q)\, .
\label{17}
\end{eqnarray}
Here  $j_1(Qr)$ is the spherical Bessel function.
The notation $\langle \nu_2\parallel A \parallel\nu_1\rangle$ 
is used for the radial integral 
$\int_0^{\infty}\d r P_{\nu_2}(r)\,A\,P_{\nu_1}(r)$.

If one neglects the partial PBrS amplitude $R_{l_2l_1}^{\pol}$ on the
right-hand side of (\ref{6}) the resulting formula coincides
with the known partial-wave expansion for OBrS 
\cite{Zhdanov1978,Tseng1989,Sobelman}.

The only characteristic in (\ref{12}) and (\ref{17}) 
which depends on the internal dynamics of the target 
is the generalized polarizability. 
However, it is exactly the calculation of this quantity which
brings a main difficulty in the case of a many-electron target,
where an accurate account for  many-electron correlations is
essential.
In 
\cite{KorolLyalinSolovyovShulakov1995,KorolLyalinSolovyovShulakov1996e,
KorolLyalinSolovyovShulakov1996c}
a simple approximate
method was introduced for the calculation of $\alpha(\om,Q)$ and
consequently of $\d\sigma_{\pol}(\om)$. 
This method allows one to avoid the rather complicated direct numerical 
computations of the many-electron correlation effects.
In short, the method can be described as follows.
Let us define a function $G(\om,q)$ equal 
to the ratio $\alpha(\om,q)/\alpha_{\d}(\om)$ of the
{\it exact} generalized and dipole polarizabilities:
\begin{equation}
\alpha(\om,q)=\alpha_{\d}(\om)\, G(\om,q)\, .
\label{8}
\end{equation}

Now let us assume that all the information about the many-electron
correlation effects is contained  in the dipole polarizability
$\alpha(\om)$, while the factor $G(\om,q)$ is not that sensitive
to them and can be calculated in the simpler approximation, for 
example within a Hartree-Fock scheme. 
Then, instead of (\ref{8}) one can write the following approximate formula:
\begin{equation}
\alpha(\om,q)
\approx
\alpha_{\d}(\om)\,
{\alpha^{\rm HF}(\om,q) \over \alpha^{\rm HF}(\om)}
\equiv   \alpha_{\d}(\om)\, G^{\rm HF}(\om,q)\,.
\label{9}
\end{equation}
This relation is the key to the method. 
The approximate equality in (\ref{9}) reduces a complex problem 
of the exact computation of $\alpha(\om,q)$  to a much simpler one: 
the calculation of the factor $G^{\rm HF}(\om,q)$ 
in the Hartree-Fock approximation.
In the papers \cite{KorolLyalinSolovyov1996,
KorolLyalinSolovyovShulakov1995,KorolLyalinSolovyovShulakov1996c,
KorolLyalinSolovyovShulakov1996e,KorolLyalinSolovyov1995a} 
the validity of this method was checked against
more rigorous calculations (carried out within various RPA-based schemes)
of $\alpha(\om,q)$ and $\d\sigma_{\pol}(\om)$ for electron scattering
on Ba, La, Eu. 
The BrS spectrum was calculated in the vicinity of the 4d-subshells 
ionization
potentials where the polarizational mechanism leads to the powerful maximum
in the spectrum.
Recently this approach was used in \cite{Romanikhin2002} to calculate,
in a broad range of photon energies, the total BrS spectra in 
$e^{-}$-Kr and $e^{-}$-La collisions.
In the cited paper the function $G(\om,q)$ (more exactly, its
inverse Fourier image, $G(\om,r)$) was calculated,  
following \cite{AstapenkoBureevaLisitsa2000}, within the local
spin density approach.
 
Another effective method for the approximate calculation of $\alpha(\om,q)$
in the $\om$-range close to the ionization thresholds of the K- and L-shells
of many-electron atoms was introduced in \cite{KorolObolenskySolovyov1998b}.
It is based on the use of the Coulomb Green function and the hydrogen-like 
wave functions for the inner shell electrons. 
In more detail this method is described in section \ref{chapter5.2}.

As was mentioned in section \ref{MainFeatures} the computation of the dynamic
response of the target is simplified considerably if one uses the 
`stripping' approximation.
The latter within the DPWA scheme can be introduced as follows
\cite{KorolLyalinSolovyovAvdoninaPratt2002}.
Using the arguments which lead to (\ref{1.1a}), one
divides the atomic electrons into two groups, the `inner' and the 
`outer' electrons, following the rule
$I_{\rm out} < \om < I_{\rm in}$.
Assuming the strong inequality $\om \ll I_{\rin}$,
one can neglect the contribution of the virtual excitations of the 
inner electrons to the sum in (\ref{4}).
Then, the amplitude $f_{\pol}$ can be approximated by the contribution of 
the outer-shell electrons alone:
\begin{eqnarray}
\fl
f^{\pol} 
\approx
-
\sum_{a,\ap=1}^{N_{\out}}\sum_{n} 
\left[
{\langle 0| {\bfe\bfr_a}| n \rangle 
\langle\bfp_2^{(-)}\, n | v_{\ap}| \bfp_1^{(+)}\, 0\rangle
\over \om - \om_{n0}  + \i 0 }
 -
{\langle\bfp_2^{(-)}\, 0 | v_{\ap}| \bfp_1^{(+)}\, n\rangle
\langle n| \bfe\bfr_a | 0 \rangle
\over \om + \om_{n0}  }
\right]\,,
\label{4a}
\end{eqnarray}
where $N_{\out}$ is the number of such electrons, and
$v_{\ap}=|\bfr-\bfr_{\ap}|^{-1}$.
 
Assuming the strong inequality $\om \gg I_{\out}$ as well, 
one expands the denominators in powers of the small parameter
$\om_{n0}/\om \sim I_{\out}/\om $ and
evaluates the leading term, proportional to $\om^{-2}$, with the 
help of closure $\sum_n \left.|n\right\rangle \left\langle n|\right. = 1$: 
\begin{eqnarray}
f_{\pol} 
\approx
{1 \over \om^2} 
\langle \bfp_2^{(-)}|\bfe\bfa_{\out}|\bfp_1^{(+)}\rangle\, , 
\label{strip.3}
\end{eqnarray}
where $\bfa_{\out}$ is the acceleration  due to the static field of the
outer electrons.

To obtain the final expression for $f_{\tot}$ let us introduce
the operator of the total acceleration $\bfa$ of 
the electron in the field of the atom,
$\bfa = -Z \bfr/r^3 + \bfa _{\rin}  + \bfa_{\out}$,
where $\bfa_{\rin}$ is
the acceleration due to the potential created by the inner electrons.
With the help of the relation between the dipole matrix elements in the
`length' and `acceleration' forms (see, e.g., \cite{Sobelman}),
the OBrS amplitude (\ref{3}) can be cast in the form
$f_{\ord}=-\om^{-2} 
\left\langle \bfp_2^{(-)}\left |\bfe\bfa\right|\bfp_2^{(+)}\right\rangle$.
Taking into account equation (\ref{strip.3}), one obtains
the following approximate formula for the total amplitude,
\begin{eqnarray}
f_{\tot}
\approx
-{1 \over \om^2} 
\left\langle\bfp_2^{(-)}
\left|\bfe\bfa_{\rm eff}\right|
\bfp_1^{(+)}\right\rangle \, ,
\label{strip.6}
\end{eqnarray}
where
$ \bfa_{\rm eff} = -Z \,\bfr/ r^3 + \bfa_{\rin}$
is the effective acceleration.

Substituting in (\ref{strip.6}) the distorted waves 
$|{\bfp}_{1,2}^{(\pm)}\rangle$ with the plane waves
$\exp(i \bfp_{1,2}\bfr)$ one derives the formula (\ref{1.1a}) for
$f_{\tot}$ within the framework of the plane-wave Born 
and the `stripping' approximations.

%%%%%%%%%%%%%% stripAr.eps
\begin{figure}
\begin{center}
\includegraphics[width=12cm,height=10cm,angle=0]{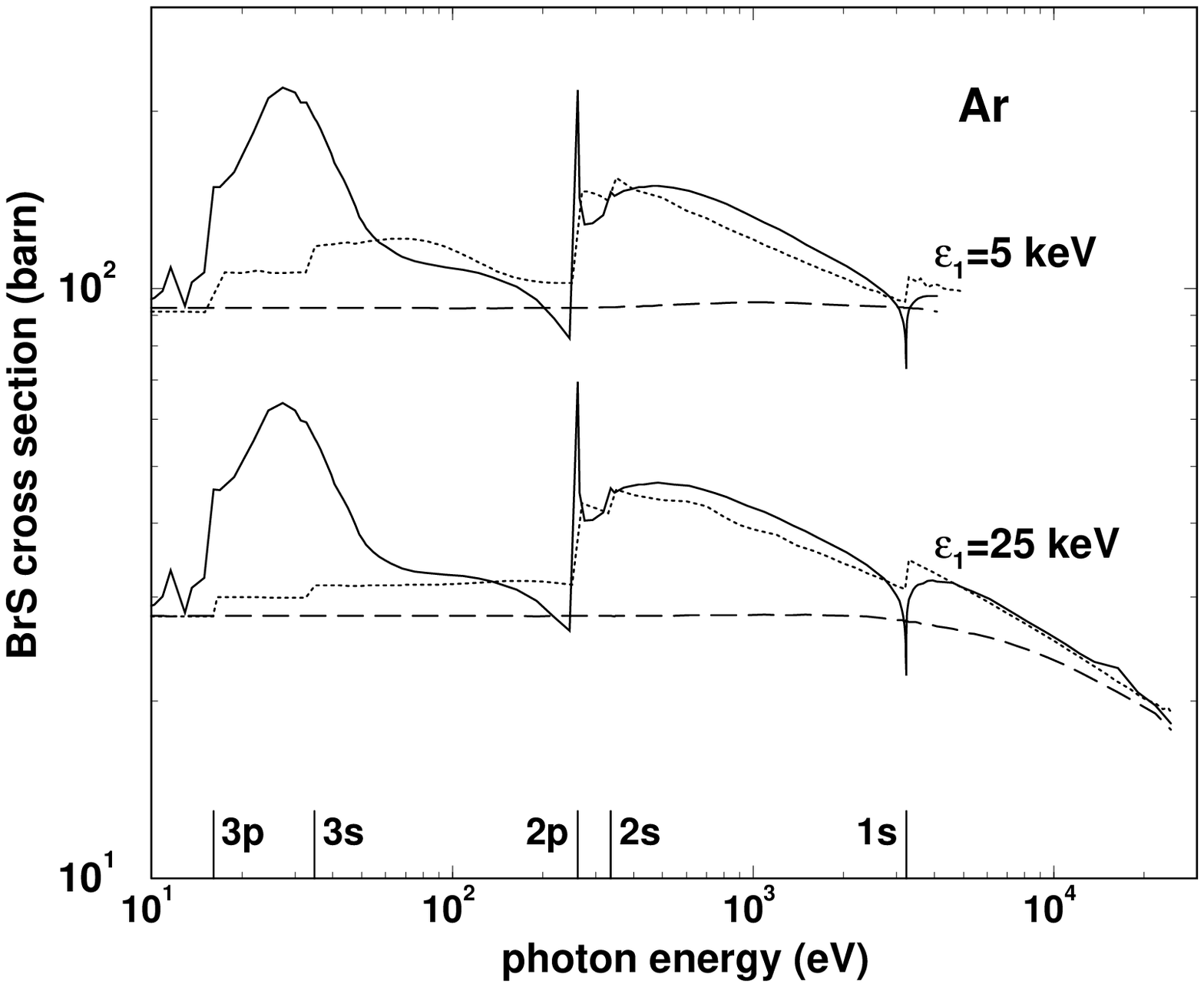}%{stripAr.eps}
\end{center}
\caption{BrS spectra, $\d \sigma(\om)$,
formed in collision of $\E_1=5$ keV and $\E_1=25$ keV
electrons with an Ar atom. 
Solid curves represent $\d \sigma_{\tot}(\om)$ within the DPWA 
and with $\alpha(\om,q)$ calculated within the RPAE
\protect\cite{KorolLyalinSolovyov1997}. 
Dashed curves describe $\d \sigma_{\ord}(\om)$,
the dotted lines correspond to the 'stripping' approximation 
(\protect\ref{strip.6}) \protect\cite{KorolLyalinSolovyovAvdoninaPratt2002}.
Vertical lines mark the non-relativistic Hartree-Fock ionization potentials 
of the atomic subshells. 
See also explanations in the text.} 
\label{figure.stripAr}
\end{figure}
%%%%%%%%%%%%%%%%%%%%%%%%%%%%%%%%%%%%%%%%%%%%%%%%%%%%%%%%%%%%%%%%%%%%%%%%%%%

\begin{figure}
\begin{center}
\includegraphics[width=12cm,height=10cm,angle=0]{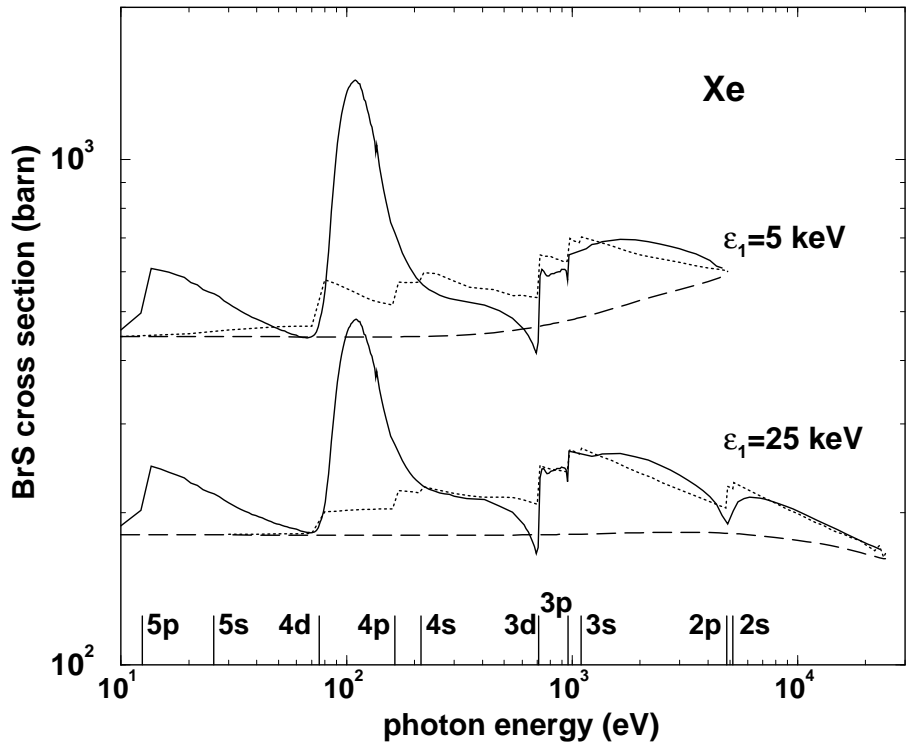}%{stripXe.eps}
\end{center}
\caption{Same as in figure \protect\ref{figure.stripAr} 
but for a Xe atom.}
\label{figure.stripXe}
\end{figure}

Figures \ref{figure.stripAr} and \ref{figure.stripXe}, where the data
on the BrS cross sections are presented for 5 and 25 keV electrons
scattered on Ar and Xe atoms, illustrate the statements made above.

It is clearly seen that  the polarizational mechanism
plays an important role in  the formation of the total BrS spectrum.
Instead of smooth curves, typical for $\d\sigma_{\ord}(\om)$ (the dashed
lines), the total BrS curves (the solid lines) exhibit complicated 
 dependence on $\om$ which is characterized  by wide powerful maxima 
and narrow cusps in the vicinity of the ionization thresholds.
Such a behavior is totally due to the contribution of the polarizational,
$\d\sigma_{\pol}(\om)$, as well as the interference,
$\d\sigma_{\rint}(\om)$, terms to the total BrS spectrum.

The total cross section was obtained from  eqs. (\ref{6})-(\ref{17}).
The calculate the generalized polarizability $\alpha(\om,q)$ for each 
value of $\om$ and $q$  we used the random-phase approximation with exchange 
\cite{AmusiaPhoto} accounting for the virtual excitations from all atomic
subshells. 
In this sense we may say that the solid lines represent the 
the `exact' $\d\sigma_{\tot}(\om)$. 

The powerful maxima in vicinity of the 3p and 3s subshells in the case of 
Ar, $\om=20\dots40$ eV, and the 4d and 4p subshells 
for Xe, $\om=80\dots110$ eV, appear mainly due to the contribution 
of the dipole excitations from these subshells to $\alpha(\om,q)$ 
and have essentially collective nature.
(This is also true for $\om \sim I_{5p}, I_{5s}$ in figure 
\ref{figure.stripXe}
but in this case the maxima are much less pronounced.)
The most part of the intensity radiated in these maxima
comes from the imaginary part of $\alpha(\om,q)$, i.e. is directly related 
to the corresponding maxima in the photoionization cross section 
(see the discussion in the paragraph containing eq. (\ref{11})).
This conclusion becomes more evident if one compares, in these $\om$ ranges,
the `exact' BrS  spectra with those obtained within the `stripping' 
approximation, the dotted lines.
Within the framework of the `stripping' approximation, eqs. (\ref{4a}) and 
(\ref{strip.3}), the imaginary part of the PBrS amplitude is ignored.
Meanwhile, it is exactly the imaginary part of $f_{\pol}$ which is related 
to the cross section of photoionization (see (\ref{12}) and then eqs. 
(\ref{1.2}) and (\ref{11})).
Therefore, the discrepancy between the solid and the dotted curves in
the vicinity of maxima is largely due to the contribution of 
${\rm Im}\Bigl(f_{\pol}\Bigr)$.
 
%%%%%%%%%%%%%% stripAr.eps
%%%%%%%%%%%%%%%%%%%%%%%%%%%%%%%%%%%%%%%%%%%%%%%%%%%%%%%%%%%%%%%%%%%%%%%%%%%

Apart from the resonant regions the `stripping' approximation 
reproduces quite well the behaviour of the `exact' curves.
The saw-like character of the dotted curves at the thresholds
clearly illustrates the physics which is behind this approximation.
Namely, for $\om$ larger than an ionization potential $I$ the total
BrS cross section increases, because the atom is `stripped' by 
a total number $N$ of the electrons in the subshell, and, therefore,
the field acting on the projectile is stronger than in the case $\om < I$.
In the formal terms the increase in the cross section is due to the
change of the sign of the interference term $\d\sigma_{\rint}(\om)$ as
$\om$ passes through the threshold: $\d\sigma_{\rint}(\om)$ is negative
for $\om < I$ and positive beyond the threshold.
These arguments explain also the additional asymmetry of the giant
resonances in the BrS spectrum and that in the spectrum of 
photoionization: in the former case the positiveness of the term
$\d\sigma_{\rint}(\om)$ slows down the decrease of the peak.

\begin{figure}
\begin{center}
\includegraphics[scale=0.7]{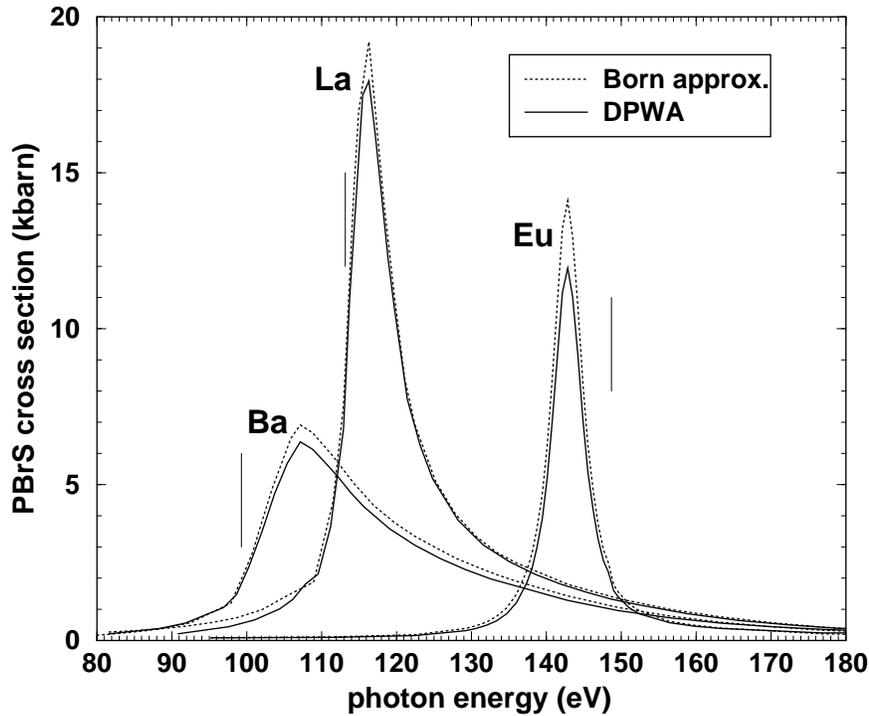}%{BaLaEu.eps}
\end{center}
\caption{Polarizational BrS cross section $\d\sigma_{\pol}(\om)$
in the vicinity of the 4d ionization potentials (marked with
vertical lines).
The incident electron energy $\E_1=250$ eV 
\protect\cite{KorolLyalinSolovyov1996a}.}
\label{figure.BaLaEu}
\end{figure}

In figure \ref{figure.BaLaEu} the dependences $\d\sigma_{\pol}(\om)$
are presented for $\om$ lying in the region of the giant resonances 
associated with the excitations from the 4d subshells in Ba, La and Eu.
The incoming electron energy is $\E_1=250$ eV.

This figure illustrates two main features of the polarizational BrS.
The first one, already mentioned, is the close relationship between
the giant resonances in $\d\sigma_{\pol}(\om)$ and those in 
$\sigma_{\gamma}$.
We note, however, that in general case, the quantity 
$\sigma_{\gamma}$  associated with the photoabsorption process 
rather than with the photoionization one alone. 
The former accounts not only for the ionization
into the continuum but the discrete excitations as well.
In those cases when excitations into the continuum dominate in the 
photoabsorption spectrum the maximum of $\sigma_{\gamma}$ lies above
the ionization threshold and so does the maximum of $\d\sigma_{\pol}(\om)$.
In connection with figure \ref{figure.BaLaEu} this is true for 
Ba and La, and does not hold in the case of a Eu atom.
For the latter is is known \cite{BeckerEu1986} that the main oscillator
strength of the 4d subshell is associated with the discrete transition
4d$\to$4f.
Therefore, the maxima of $\sigma_{\gamma}$ and $\d\sigma_{\pol}(\om)$
are located below $I_{4d}$.

%%%%%%%%%%%%%% BaLaEu.eps
%%%%%%%%%%%%%%%%%%%%%%%%%%%%%%%%%%%%%%%%%%%%%%%%%%%%%%%%%%%%%%%%%%%%%%%%%%%

The solid lines in figure \ref{figure.BaLaEu} were obtained within 
the DPWA scheme with the partial amplitudes of PBrS computed from
(\ref{17}).
The dashed lines correspond to $f_{\pol}$ within the Born approximation
(see eq. (\ref {1.1})).
In both cases to calculate the generalized dynamic polarizability 
we used the RPAE with relaxation scheme \cite{AmusiaPhoto} for Ba and La, 
and  the spin-polarized RPAE \cite{AmusiaDolmatovIvanov1983} for Eu. 

It is seen that for incident energies as low as 250 eV the Born
approximation gives almost the same result as the DPWA. 
The reason for this coincidence lies in the fact that, 
in contrast to the process of ordinary BrS, polarizational radiation 
is formed mainly at large distances, $r\sim p_1/\om$, between 
a projectile and an atom \cite{Zon1979,AmusiaZimkinaKuchiev1982}, where 
the distorting influence of the atomic potential on the projectile's 
movement is comparatively small.
Hence, to calculate the polarizational component of the BrS spectrum, 
one may use the Born approximation, which results in the formula
\begin{equation}
\d\sigma_{\pol}^{\rm B}(\om) 
  =
{16 \over 3}\, 
{\om^4 \over c^3 p_1^2}
\int \limits_{q_{\min}}^{q_{\max}} {\d q \over q}\,
\left|\alpha(\om,q)\right|^2\, ,
\label{7}
\end{equation}
where $q_{\min}=p_1-p_2$, $q_{\max}=p_1+p_2$.
From a computational viewpoint this expression can be
evaluated with much less efforts than its analogue within the DPWA
scheme.

The further step in simplifying the theoretical analysis of the 
polarizational part of the spectrum is to use the approximation 
(\ref{9}) for the generalized polarizability.
Then, instead of (\ref{7}), one arrives at
\begin{equation}
\d\sigma_{\pol}^{\rm B}(\om) 
\approx
{16 \over 3}\, {\om^4 \over c^3 p_1^2}\,
 \left|\alpha_{\d}(\om)\right|^2
\int \limits_{q_{\min}}^{q_{\max}} {\d q \over q}\,
\left|G^{\rm HF}(\om,q)\right|^2\, .
\label{10}
\end{equation}

\begin{figure}
\begin{center}
\includegraphics[scale=0.7]{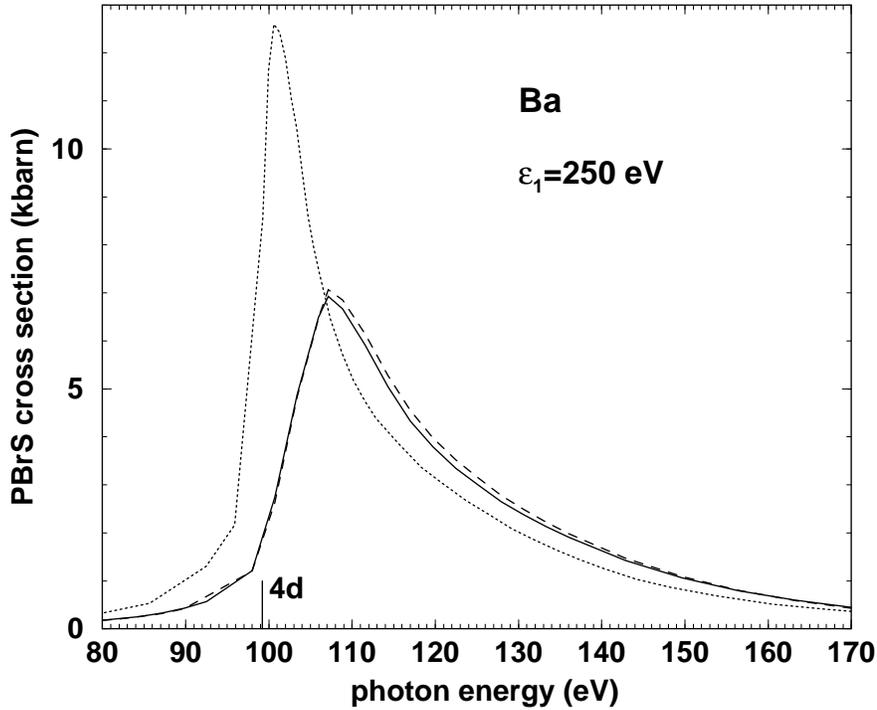}%{BaPol.eps}
\end{center}
\caption{Polarizational BrS cross section $\d\sigma_{\pol}(\om)$
for $e^{-}+$Ba in the vicinity of the 4d ionization potential (marked with
the vertical line). 
The solid curve was 
calculated using the exact Born formula (\protect\ref{7}),
the dashed curve corresponds to (\protect\ref{10}).
The dotted line originates from (\protect\ref{7}) but with
$\alpha(\om,q)$ within the Hartree-Fock approximation.
\protect\cite{KorolLyalinSolovyov1996,KorolLyalinSolovyovShulakov1995}}
\label{figure.BaPol}
\end{figure}

To check the validity of the proposed method in 
\cite{KorolLyalinSolovyovShulakov1995,KorolLyalinSolovyovShulakov1996b,
KorolLyalinSolovyov1996} the
polarizational part of the spectrum for was calculated for 
0.25 - 10 keV electrons on Ba, La and Eu 
using the exact Born formula (\ref{7}) and the approximate one 
(\ref{10}). 
In all cases considered the results are close.
Figure \ref{figure.BaPol} illustrates this for the collision 
of a 250 eV electron with Ba. 
The discrepancy between the solid curve, corresponding to (\ref{7}) with
$\alpha(\om,q)$ calculated within the RPAE, and 
the short-dashed one, representing (\ref{10}) where the correlations
were accounted for in the factor $\alpha_{\d}(\om)$ only, 
is almost negligible. 
For the sake of comparison, we also present the polarizational
cross section calculated via (\ref{7}) but with $\alpha(\om,q)$ obtained
in the Hartree-Fock approximation 
(dotted curve).

%%%%%%%%%%%%%%BaPol.eps
%%%%%%%%%%%%%%%%%%%%%%%%%%%%%%%%%%%%%%%%%%%%%%%%%%%%%%%%%%%%%%%%%%%%%%%%%%%

Another possibility to deduce $\alpha(\om,q)$ is to combine 
(\ref{9}) with (\ref{11}).
Provided the dependence $\sigma_{\gamma}(\om)$ is known over a sufficiently
wide $\om$-region one can restore the real part of $\alpha_{\d}(\om)$ from
the dispersion relation. 
Such a way of obtaining $\alpha(\om,q)$  is especially useful when 
direct calculations are difficult to perform 
(for example, when the BrS process is investigated in a dense
media rather that in a pure `one electron -- one atom' collision).

\begin{figure}
\begin{center}
\includegraphics[scale=0.7]{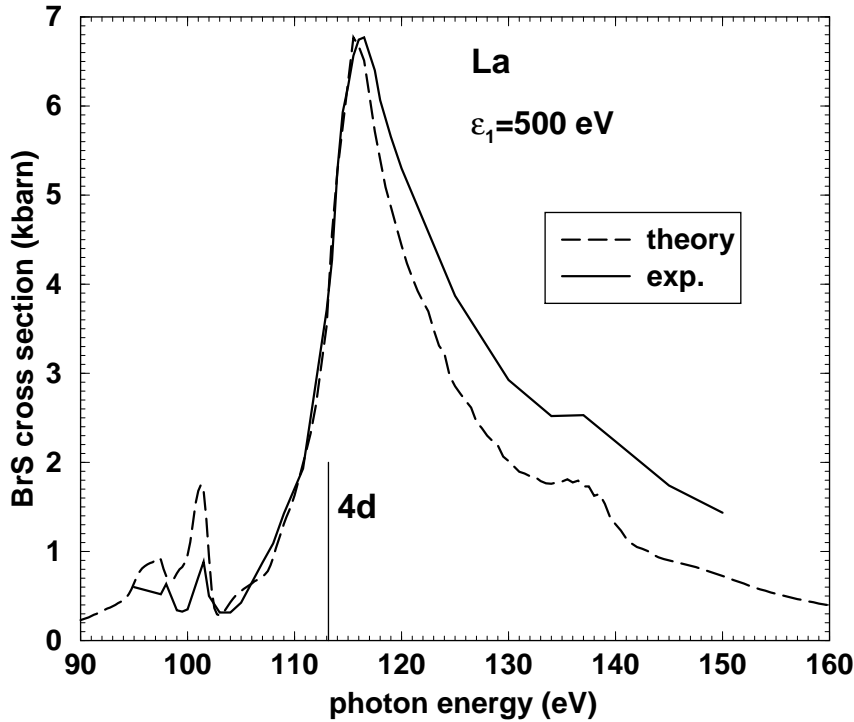}%{La500a.eps}
\end{center}
\caption{Comparison of experimentally measured
 (dashed curve) \protect\cite{ZimkinaShulakovBrajkoEtAl1984a}
BrS cross section formed in the collision of 500 eV electrons with La
with the calculated dependence (solid curve)
\protect
\cite{KorolLyalinSolovyovShulakov1996c,KorolLyalinSolovyovShulakov1996b}.
See also the explanation in the text.
}
\label{figure.La500}
\end{figure}

To illustrate this approach in figure \ref{figure.La500} 
the experimentally measured emission spectra for 500 eV electrons
on La \cite{ZimkinaShulakovBrajkoEtAl1984a} are compared with the
theoretical results 
\cite{KorolLyalinSolovyovShulakov1996c,KorolLyalinSolovyovShulakov1996b}.
The experiment was carried out with the metallic La, therefore,
the use of $\alpha(\om,q)$ (or $\alpha_{\d}(\om)$) calculated for
an isolated La atom seems not a fully adequate approach.
To avoid this problem in \cite{KorolLyalinSolovyovShulakov1996c} 
the experimental data for the photoabsorption spectrum 
\cite{ZimkinaGribovski1971,HenkeGulliksonDavis1993} was used  
to calculate the dynamic dipole polarizability, 
which then was substituted into (\ref{10}).  
The factor $G^{\rm HF}(\om,q)$ was calculated in the Hartree-Fock 
approximation.
The experimental data  \cite{ZimkinaShulakovBrajkoEtAl1984a} have no 
absolute scale, so it was normalized to the magnitude of theoretical curve 
at the maximum. 
The background radiation, i.e. the ordinary BrS, was 
subtracted from the measured spectrum. 
Therefore, the experimental curve in the figure represents by itself a 
sum of the polarizational and the interference terms.

%%%%%%%%%%%%%% La500.eps
%%%%%%%%%%%%%%%%%%%%%%%%%%%%%%%%%%%%%%%%%%%%%%%%%%%%%%%%%%%%%%%%%%%%%%%%%%%

The agreement between the two curves is  quite good.  
The main discrepancy is seen on the right wing of the spectrum 
where the experimental curve lies higher than the calculated one. 
This discrepancy may be attributed to the contribution of the 
interference term which was omitted in the calculations.

%%%%%%%%%%%%%%%%%%%%%%%%%%%%%%%%%%%%%%%%%%%%%%%%%%%%%%%
\section{BrS in collisions of heavy particles.\label{chapter5.2}}

In this section we present the formalism and 
the results of numerical calculations of the BrS spectra
formed in a collision of two heavy particles.
We restrict ourselves to the case of `fast' collisions, when
the translational motion of the colliders can be described by
 plane waves. 
The peculiar features of the BrS formed in slow collisions, and,
in particular, its relationship to the well-known molecular orbital
radiation, are discussed in \cite{Solovyov1992}.

Let us describe in short the formalism which
allows one to calculate the BrS spectrum formed in fast non-relativistic
collisions of two atomic particles.
Following the papers  
\cite{AmusiaKuchievSolovyov1984,AmusiaKuchievSolovyov1985},
where such a problem was solved for the first time,
we consider the general case when both  colliders, a projectile 
and a target,  have internal dynamic structure.
In what follows we use the term `atom 1' for a  projectile, and `atom 2'
for a target.

Let $M_{1,2} \gg 1$ denote the masses of the atoms, $Z_{1,2}$ stand
the charges of the nuclei, ${\bf p}_{1,2}$ and 
${\bf p}_{1,2}^{\prime}$ notate 
the translation momenta before and after the collision, respectively.
Apart from the translation momentum the state of 
each particle is characterized by the set of quantum numbers which
refer to its internal (electronic) state.
We assume that both atoms before and after the collision are in their
grounds states which are notated with '0'.
The interaction between the atoms is described by the potential 
$\hat{U}\equiv U(\{\bfr\}_1,\{\bfr\}_2)$
(here $\{\bfr\}_j$, $j=1,2$, denote the coordinates of all the particles
in atom '1' and atom '2'), which is the sum of the pair interactions 
between the constituents of the two atoms.
For each one of the colliders the interaction with the field of 
radiation is described by the operator $V$ which has the general form
$\hat{V} = \sum_{i=1}^{N} (e_i/m_i)\exp(-\i\bfk\bfr_i)\, (\bfe\hat{\bfp}_i)$.
Here the sum is carried out over the atom's constituents of the total
number $N$, the quantities $e_i$ and $m_i$ stand for the charges and
the masses of the constituents, and $\hat{\bfp}_i$ is the momentum 
operator.

\begin{figure}
\begin{center}
\includegraphics[scale=0.75]{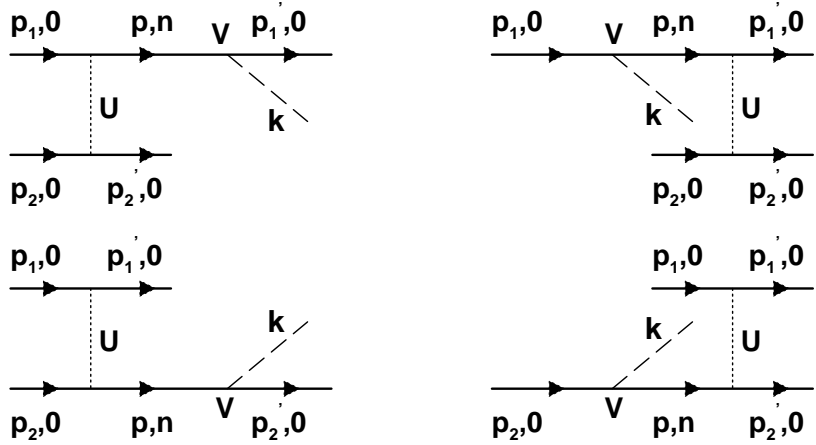}%{Diagr_H.eps}
\end{center}
\caption{Diagrammatical representation of the BrS amplitude 
in an atom-atom  collision.}
\label{fig:diagratat}
\end{figure}

In the lowest orders of the perturbation theory in $U$ and $V$ 
the process of photon emission is represented by four diagrams 
drawn in figure   (\ref{fig:diagratat}).
The solid lines denote the atoms, the dashed lines stand for the emitted
photon, whose momentum is $\bfk$, energy $\om$ and polarization $\bfe$,
the vertical dotted lines represent the interaction $\hat{U}$.
The two upper diagrammes correspond to the photon emission by a 
projectile which is virtually polarized by the target.
The amplitude of this process is written as follows:
\begin{eqnarray}
f_1
=
\int {\d \bfp \over (2\pi)^3} \sum_n
&\Biggl[
{
\langle \bfp^{\prime}_1, 0\left|\hat{V}\right|\bfp;n \rangle
\langle\bfp,n;\,\bfp_2^{\prime},0\left|\hat{U}\right|\bfp_1,0;\,\bfp_2,0\rangle
\over 
\E_n-\E_0 + E_{\bfp} - E_{\bfp_1} + E_{\bfp_2^{\prime}} -E_{\bfp_2}
}
\nonumber\\
&
+
{
\langle\bfp_1^{\prime},0;\,\bfp_2^{\prime},0\left|\hat{U}\right
|\bfp,n;\,\bfp_2,0\rangle
\langle \bfp, n\left|\hat{V}\right|\bfp_1;0 \rangle
\over 
\E_n-\E_0 + \om + E_{\bfp} - E_{\bfp_1}
}
\Biggr]\,.
\label{ampl.f1}
\end{eqnarray}
Here $E_{\bfp}=p^2/2M$ stands for the kinetic energy of the 
translation motion,
$\E$ denotes the energy of the electronic state of the projectile,
the integration is carried out over the translation momentum of the
projectile in the intermediate (virtual) state, and the sum $\sum_n$ 
is evaluated over the whole spectrum (including the excitations into the 
continuum) of the internal degrees of freedom of the atom 1. 

%%%%%%%%%%%%%%%%%%%%%%%%%%%%%%%%%

The process of the photon emission by the target, is  described by the 
lower pair of diagrammes, and defines the amplitude $f_2$.
The analytic expression for $f_2$ one obtains from (\ref{ampl.f1})
by exchanging the characteristics of the atom `1' with those of the atom `2'.

The total BrS amplitude is given by
\begin{equation}
f = f_1 + f_2.
\label{v89p1512n8}
\end{equation}

To evaluate the terms $f_{1,2}$ one starts  
with the factorization of the center-of-mass motion in the definition
of the wavefunction of a complex system: 
%%%%%%%% 5.1 %%%%%%%%%%%%%%%%%%%%%%%%%%%%
\begin{equation}
\Psi_{\bfp n}(\{\bfr_i\}
=
\ee^{\i \bfp \bfR}
\psi_n \left(\{\bfr_i-\bfR\}\right),
\label{v89p1512n1}
\end{equation}
where $\bfp$ is the momentum of the center-of-mass,
$\bfR$ is its radius-vector,
$\psi_m\left(\{\bfr_i-\bfR\}\right)$ 
is the wavefunction describing the internal degrees of freedom of the 
system's constituents with the subscript `$n$' staying for all 
the necessary quantum  numbers.

Then, integrating over the motion of the center-of-masses
of the colliders, one expresses the matrix elements
$\langle \bfp^{\prime}, m^{\prime}\left|\hat{V}\right|\bfp;m \rangle$ and
$\langle \bfp_2^{\prime}, m_2^{\prime}; \bfp_1^{\prime}, m_1^{\prime}
\left|\hat{U} \right| \bfp_1, m_1 ; \bfp_2, m_2 \rangle$
 in terms of the matrix elements between the corresponding
wavefunctions $\psi$ which describe the internal states.
Finally, accounting for $v_{1,2}/c \ll 1$ ($v_{1,2}$ are the velocities 
of the colliders) and carrying out the dipole-photon limit
$\om (R_{\rm at})_{1,2 }/c \ll 1$  
one arrives at:
\begin{eqnarray}
f_1
=
{4\pi(\bfe\bfq)\over q^2}\,
\Bigl(Z_2-F_2(q)\Bigr)\,
A_1(\om,q)\,,
\label{v89p1512n10_a}
\end{eqnarray}
where $\bfq=\bfp_1-\bfp_1^{\prime}=\bfp_2^{\prime}-\bfp_2$ is the momentum 
transfer.
The function $A_1(\om,q)$ is defined as follows:
\begin{equation}
A_1(\om,q)
=
{Z_1-F_1(q) \over M_1 \om}\, e_1 - \om\alpha_1(\om,q)\, ,
\label{v89p1512n10}
\end{equation}
where $\alpha_1(\om,q)$ is the generalized polarizability of the projectile,
and $e_1$ is its net charge.
In (\ref{v89p1512n10_a}) and (\ref{v89p1512n10}) the quantities
$F_{1,2}(q)$ are the form-factors of the atoms.

Interchanging in (\ref{v89p1512n10_a}) the indices 1 and 2 
one obtains the amplitude $f_2$.
Therefore, the total amplitude of the BrS in atom-atom scattering
is
\begin{eqnarray}
f
=
{4\pi(\bfe\bfq)\over q^2}\,
\Biggl[
\Bigl(Z_2-F_2(q)\Bigr)\,A_1(\om,q)
-
\Bigl(Z_1-F_1(q)\Bigr)\,A_2(\om,q)
\Biggr]
\label{v89p1512n12}
\end{eqnarray}

This general formula generalizes the result of the Born approximation
in application to the structureless charged particle scattering on 
a many-electron target,   see (\ref{1.1}).
The right-hand side of (\ref{v89p1512n12}) reduces to that of  (\ref{1.1})
if one considers $\alpha_1(\om,q)=0$ and $F_1(\om,q)=0$.
In this limit the first term in (\ref{v89p1512n12}) reduces to $f_{\ord}$,
whereas the second one reproduces the PBrS amplitude.

Another feature which we would like to note in connection with 
(\ref{v89p1512n12}) is the vanishing of the total BrS amplitude 
in the collision of two identical particles. 
Indeed, in this case the moduli of both terms are equal and the minus
sign leads to $f=0$.
This is not surprising if one recalls that (\ref{v89p1512n12}) corresponds
to the dipole-photon approximation, in whose framework any system of 
identical particles (no matter how complex they are) does not radiate.
This restriction does not hold beyond the dipole approximation.
The most rigorous treatment of the BrS problem in the non-dipolar 
domain should also include the consideration of the relativistic effects.
The formalism describing the BrS radiation in relativistic collisions
of atomic particles can be found in \cite{AmusiaSolovyov1990a}.

Using the amplitude (\ref{v89p1512n12}) one derives the following
expression for the spectral distribution of the BrS radiation in 
atom-atom collisions:
%%%%%%%% 5.17 (16?) %%%%%%%%%%%%%%%%%%%%%%%%%%%%
\begin{equation}
{\d\sigma\over \d\om}
=
{16 \om^3\over 3c^3 v_1^2}
\int_{q_{\min}}^{q_{\max}}
{\d q \over q}\,
\Biggl|
\Bigl(Z_2-F_2(q)\Bigr)\,A_1(\om,q)
-
\Bigl(Z_1-F_1(q)\Bigr)\,A_2(\om,q)
\Biggr|^2\,.
\label{v89p1512n15}
\end{equation}
Here $q_{\min}$ and $q_{\min}$ are the minimum and maximum momentum
transfer:
%%%%%%%% 5.26 %%%%%%%%%%%%%%%%%%%%%%%%%%%%
\begin{eqnarray}
\fl
\quad
q_{\min} = \mu v_1 \left(1-\sqrt{1-{2\om / \mu v_1^2}}\right),
\qquad
q_{\max}  =  \mu v_1 \left(1+\sqrt{1-{2\om / \mu v_1^2}}\right)\,,
\label{QmaxQmin}
\end{eqnarray}
and $\mu$ stands for the reduced mass of the colliders.

%%%%%%%%%%%%%%%%% alpha+Xe
\begin{figure}%[t]
\begin{center}
\includegraphics[width=12cm,height=9cm,angle=0]{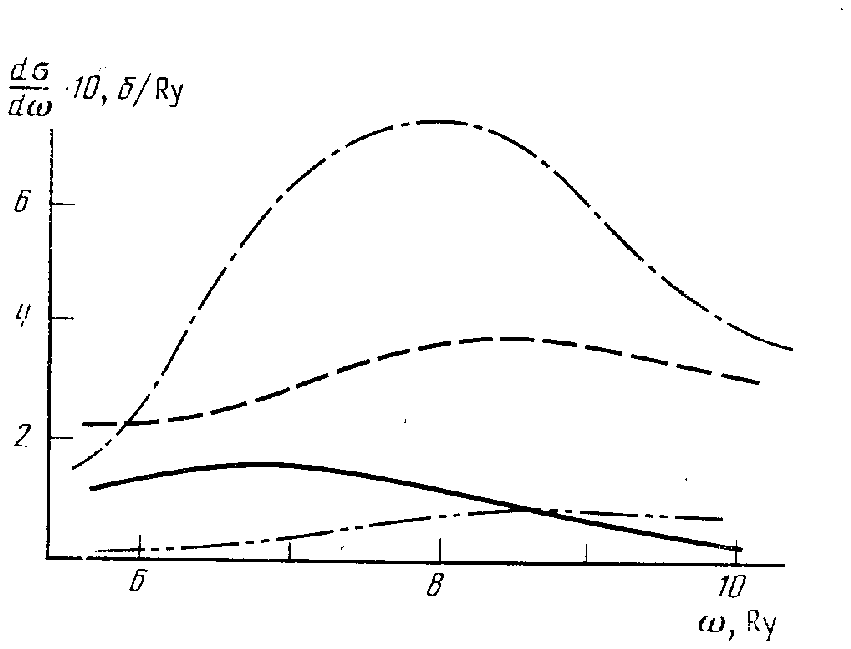}%{xe-a.eps}
\end{center}
\caption{BrS spectral distribution (in barn/ryd)  
in collisions of various particles with a Xe atom
\protect\cite{AmusiaKuchievSolovyov1985c}.
Dashed line represents the total BrS spectrum for an electron,
dashed-dotted line  stands for $\d\sigma/\d \om$ for an $\alpha$-particle,
dashed-double-dotted line - the spectrum for a He atom with
no account for $\alpha_{\rm He}(\om,q)$, solid line describes the spectrum
for He+Xe collision with account for  $\alpha_{\rm He}(\om,q)$.
In all cases the initial velocity of the projectile is 5 a.u.
}
\label{fig:HeXe}
\end{figure}

It was mentioned in section \ref{MainFeatures} that for the same 
initial velocity of the intensity of the BrS formed in 
the collisions of a heavy particle with a many-electron atom 
can be comparable or even higher that that in the electron-atom 
collision.
To illustrate this we included figure \ref{fig:HeXe} 
where the dependences $\d \sigma/\d \om$ are presented for the collisions
$e^{-}+$Xe, $\alpha+$Xe and He$+$Xe for the range of photon energies in 
the vicinity of the ionization potential of the 4d subshell in Xe, 
$I_{4d}=73.4$ eV. 
All curves correspond to the initial velocity $v_1=5$ a.u.

The spectra were calculated with the help of (\ref{v89p1512n15}) 
the integrand
of which acquires simplified forms (as was mentioned above) in the case
of electron- and $\alpha$-Xe collisions.
The form-factors of He and Xe were calculated in the Hartree-Fock 
approximation,
the generalized polarizability of Xe, $\alpha_{\rm Xe}(\om,q)$ 
within the RPAE scheme as it was done in 
\cite{AmusiaAvdoninaChernyshevaKuchiev1985}.
The generalized polarizability of He was approximated by 
$\alpha_{\rm He}(\om,q) \approx -F_{\rm He}(q)/\om^2$ (cf. (\ref{1.4})) 
since in 
the considered range of the photon energies ($6\dots 10$ ryd) 
the strong inequality $\om \gg I_{He}\approx 2$ ryd is valid.
Although 
$\Bigl|\alpha_{\rm He}(\om,q)\Bigr|^2 \ll 
\Bigl|\alpha_{\rm Xe}(\om,q)\Bigr|^2$
in the range $\om \sim I_{4d}$, the two terms in the brackets  
in the integrand from (\ref{v89p1512n15}) are of the same order of magnitude,
and the interference between them is important when calculating the spectrum.

The most striking feature clearly seen in the figure is that 
in the whole spectral interval the magnitude of $\d\sigma/\d\om$ for
the collision $\alpha+$Xe exceeds that formed in  the collision $e^{-}+$Xe.
The qualitative explanation of this effect is as follows.
In the collision $e^{-}+$Xe both channels, the ordinary and 
the polarizational,
contribute to the amplitude/spectrum. 
The amplitude of the process is described by (\ref{1.1}) where one uses 
$m=1$ and $e=1$.
In the case of $\alpha$-particle scattering, $f_{\ord}\to 0$ because of a 
large mass of the projectile, but $f_{\pol}\propto 2\alpha_{\rm Xe}(\om,q)$ 
is two times higher than for an electron.
Therefore, the polarizational part of the spectrum for an $\alpha$-particle 
is approximately four times larger than the PBrS of an electron.
In the latter case, however, there are contributions of the terms 
$\d\sigma_{\ord}$ and $\d\sigma_{\rint}$ which reduce the discrepancy.

Another important feature which is illustrated by the figure is that the 
intensity  of the radiation in collision of a  neutral and compact He atom 
with Xe is, in the order of magnitude, equal to that for $\alpha$- and 
$e^{-}$-Xe collisions.
This is totally due to the polarizational BrS which appears as a result 
of virtual polarization of Xe during the collision.
 
In the case of a heavy charged structureless projectile
the ordinary BrS can be neglected (the first terms in (\ref{1.1} and in the 
brackets in (\ref{v89p1512n15})) and the spectrum is  defined 
by the polarizational channel only. 
In this case  (\ref{v89p1512n15}) reduces to
%%%%%%%% 
\begin{equation}
\d\sigma_{\pol}(\om,v_1)
\equiv
\om\,{\d\sigma\over \d\om}
=
{16 e^2 \om^4\over 3c^3 v_1^2}
\int_{q_{\min}}^{q_{\max}}
{\d q \over q}\,
\Bigl|
\alpha(\om,q)
\Bigr|^2\,,
\label{scaling.1}
\end{equation}
where $e$ is the charge of projectile.

Various methods of calculation of the generalized polarizability 
$\alpha(\om,q)$ in the $\om$-regions in vicinities of the ionization
potentials of the atomic subshells as well as in the limit 
$\om \gg I_{1s}$ have been already discussed above in the paper.
Here we want to mention the approach, developed recently 
\cite{KorolObolenskySolovyov1998b,KorolObolenskySolovyov1999},
which is very effective for the description of the dynamic response 
in the photon energy range corresponding to the ionization
potentials of the  the K- and L- atomic shells.
In this case the electrons of these shells provide the main contribution
to the polarizational BrS cross section.
The method is based on the use of the non-relativistic Coulomb Green's
function (see, e.g. \cite{ZapryagaevManakovPalchikov1985}) and the 
hydrogen-like wave functions for the inner shell electrons calculated 
in the field of the effective nucleus charge $Z_{\eff}$.
The main difficulty in the description of the many-electron
system in terms of the single electron hydrogen-like wave functions
consists in the correct accounting for the Pauli principle.
In the cited paper it was demonstrated that the Pauli principle 
can be taken into account by using the subtraction procedure, 
which cancels the electron transitions
from the ground state to the occupied atomic levels.
Hence, within the framework of this approach the exact 
generalized polarizability $\alpha(\om,q)$ of the inner subshell 
with quantum numbers $n$ and $l$ 
(the principal and the orbital quantum numbers) is approximated 
as follows:
\begin{equation}
\alpha_{nl} (\omega,q)
\approx
\talpha_{nl} (\omega,q)
-
\sum_{\np\lp}\talpha_{nl\to \np\lp} (\omega,q)\,.
\label{scaling.2}
\end{equation}
Here $\talpha_{nl} (\omega,q)$ is the generalized dynamic polarizability
of the $nl$-state but calculated in the point Coulomb field of the 
charge $Z_{\eff}$.
The latter can be chosen, for example, from the condition
$I_{nl}= Z_{\eff}^2/2n^2$, i.e. the ionization potential is approximated 
by the corresponding hydrogen-like value,
or one may crudely put $Z_{\eff}=Z$ in the case of the K-shell electrons.
The quantities $\np$ and $\lp$ on the right-hand side of
(\ref{scaling.2}) are the quantum numbers of the occupied 
subshells in the many-electron atom the transition to which 
is allowed by the dipole selection rules but is forbidden by the 
Pauli  principle.
Hence, each term $\talpha_{nl\to \np\lp} (\omega,q)$ corresponds to the 
contribution to $\talpha_{nl\to \np\lp} (\omega,q)$ 
due to the discrete dipole transition $(n,l)\to (\np,\lp)$.
To meet the Pauli principle all these terms must be subtracted from
$\talpha_{nl} (\omega,q)$.

The total polarizability of the atom is the sum of all 
$\alpha_{nl} (\omega,q)$.
The approximation (\ref{scaling.2}) fails for the subshells which 
exhibit a strong correlated reaction to the action of an external field.
For such subshells the model of independent electrons  is inapplicable.
In contrast, the dynamics of electrons in the inner shells
(K- and/or L-shells) is mainly governed by the action of the field of  
nucleus which outpowers the inter-electron correlations. 
Therefore, the substitution of the exact polarizability 
$\alpha_{nl} (\omega,q)$ with its point Coulomb analogue is justified.
In many-electron atoms the ionization potentials of the K- and L-shells
are well-separated. 
Therefore, if one is interested in the magnitude of $\alpha(\omega,q)$
in the range, for example, $\om \sim I_K$, then one can neglect the
contribution of all the shells but K to the generalized polarizability 
and to approximate 
$\alpha(\omega,q)\approx \alpha_{1s}(\omega,q)\approx
\talpha_{1s}(\omega,q)-\sum_{\np\lp}\talpha_{1s\to \np\lp} (\omega,q)$. 
The analytic formulae for $\talpha_{nl}(\omega,q)$ and 
$\talpha_{1s\to \np\lp} (\omega,q)$ are presented in 
\cite{KorolLyalinObolenskySolovyov1998}.

%%%%%%%%%%%%%%%%%
\begin{figure}%[h]
\begin{center}
\includegraphics[width=13cm,height=11cm,angle=0]{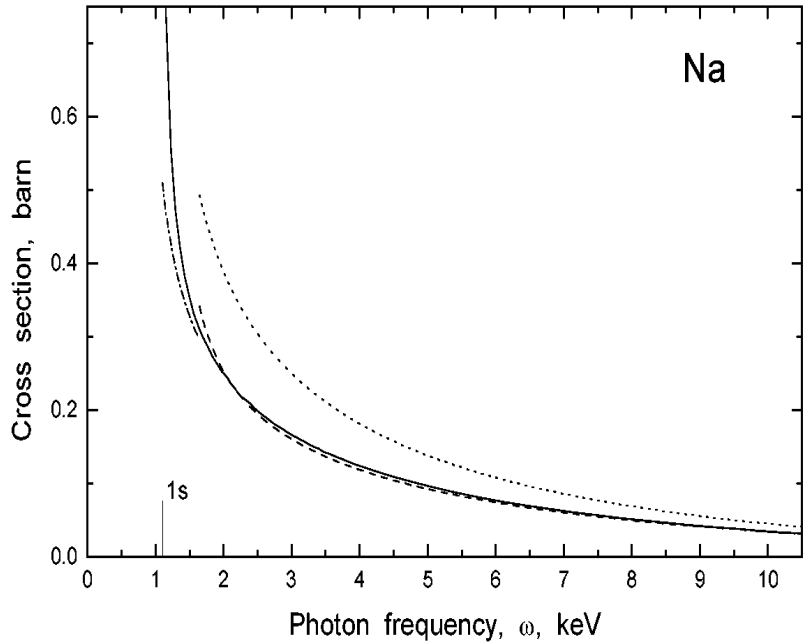}%{Na1.eps}
\end{center}
\caption{
The PBrS cross sections $\d\sigma_{\pol}(\om,v_1)$ 
(see (\protect\ref{scaling.1})) for the collision of a proton with 
a Na atom ($Z=11$) for the photon energies above the K-shell ionization 
potential (marked by a vertical line)
\protect\cite{KorolObolenskySolovyov1998b}.
The collision velocity  is $v_1= 40$ au.
Full curve, the cross section obtained in the Hartree-Fock approximation.
Dotted curve, the cross section obtained in the hydrogen-like model 
(with $Z_{\eff}=Z$)
without the Pauli principle taken into account.
Broken curve, the cross section obtained in the hydrogen-like model
($Z_{\eff}=Z$)
with the Pauli principle taken into account.
Chain curve, the cross section obtained in the hydrogen-like model 
(with $Z_{\eff}=9$)
with the Pauli principle taken into account.}
\label{scaling.fig1}
\end{figure}
\begin{figure}%[h]
\begin{center}
\includegraphics[width=12cm,height=9cm,angle=0]{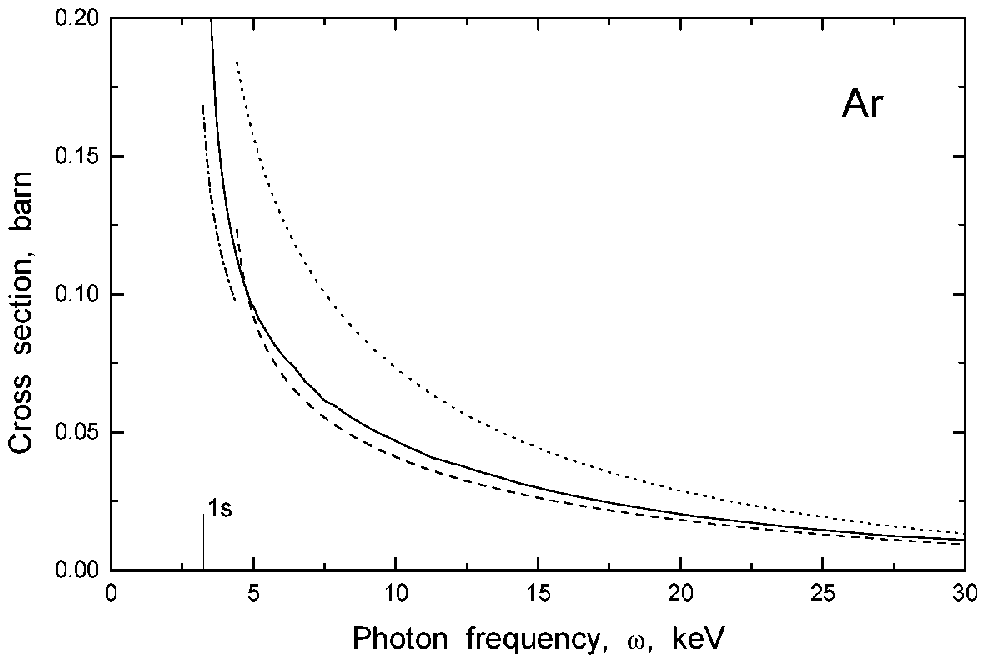}%{Ar1.eps}
\end{center}
\caption{Same as in figure \protect\ref{scaling.fig1} but for
an Ar atom ($Z=18$).
The chain curve was obtained for $Z_{\eff}=15.4$.
}
\label{scaling.fig2}
\end{figure}

%%%%%%%%%%%%%%%%%
In figures \ref{scaling.fig1} and \ref{scaling.fig2} we present the 
results of  calculations of the PBrS cross sections in vicinities
of the K-shell ionization potentials  Na and  Ar atoms.
Full curves represent the polarizational BrS cross section
(\ref{scaling.1}) $\alpha(\om,q)$ obtained in the Hartree-Fock 
approximation.  
Dotted curves represent the PBrS cross section
calculated with the use of the hydrogen-like polarizability 
$\talpha_{1s} (\omega,q)$ which does not take into account the 
Pauli principle (see \ref{scaling.2}). 
For each target the effective charge $Z_{\eff}$ was chosen from
the condition $Z_{\eff}=Z$.
The comparison of dotted and solid curves shows that
the cross sections obtained in the Hartree-Fock and 
the purely hydrogen-like approximations have the same 
order of magnitude but do not agree well enough.

%%%%%%%%%%%%%%%%%
%%%%%%%%%%%%%%%%%

The discrepancy between the two models becomes less pronounced 
if one takes into account the Pauli principle. 
The results of the calculation of the PBrS cross sections in this case
are plotted as broken curves which, as it is seen from the figures,
reproduce the behaviour of the Hartree-Fock cross sections reasonably well.  
The discrepancy between the two approaches is comparable with 
the accuracy of the Hartree-Fock method itself.
The deviation of the hydrogen approximation from the Hartree-Fock one
becomes significant in the vicinity  of the hydrogen-like ionization
potential of the K-shell. 
Shifting of the hydrogen-like ionization potential of the K-shell 
occurs due to the inter-electron interaction.  
This phenomenon can be taken into account within the frame
of the hydrogen-like model by choosing the value of $Z_{\eff}$ 
from the condition $I_{1s}= Z_{\eff}^2/2$ instead of $Z_{\eff}=Z$
($I_{1s}$ is the Hartree-Fock ionization potential of the K-shell).
The PBrS cross sections calculated with this values of $Z_{\eff}$ 
are shown in figures by chained curves.
The figures demonstrate that the imaginary continuation of the chained
curves smoothly matches the broken curves in the vicinity of the
hydrogen-like ionization potential.  
This means that the hydrogen-like approach  provides a simple and 
effective method for the PBrS cross section calculations 
applicable over the whole range of photon
energies above the ionization potentials of the inner atomic shells.

The use of the hydrogen-like approximation for the description of 
the generalized atomic polarizability allowed to evaluate the scaling
law for the polarizational BrS cross section 
\cite{KorolObolenskySolovyov1998b} which, for a heavy projectile,
defines the total BrS cross section.
In the hydrogen-like model the distance and energy can be scaled by 
the factors $1/Z_{\eff}$ and $Z_{\eff}^2$ respectively.
Therefore, one can derive the following 
scaling law for the generalized polarizability $\alpha_{nl}(\omega,q)$
defined in (\ref{scaling.2}):
\begin{equation}
\alpha_{nl}(\omega,q)
=
{ N_{nl} \over Z_{\eff}^4}\,
 \alpha^{\rm red}_{nl}
\left(
{\omega\over Z_{\eff}^2},{q \over Z_{\eff}}
\right),
\label{polscal}
\end{equation}
where $N_{nl}$ is the number of electrons in the subshell $(nl)$,
and $\alpha^{\rm red}_{nl}$ denotes the expression on the right-hand
side of (\ref{scaling.2}) calculated for the $(nl)$-subshell of
a hydrogen atom.

Using (\ref{polscal}) in (\ref{scaling.1}) one expresses the cross section
$\d\sigma_{\pol}(\om,v_1)$ calculated in the the range $\om \sim I_{nl}$
and for the projectile of velocity $v_1$ in terms of 
$\d\sigma_{\pol}^{\rm red}(\om/Z_{\eff}^2,v_1/Z_{\eff})$, which is   
the BrS cross section for the collision with a hydrogen atom but 
for the scaled velocity $v_1/Z_{\eff}$  and the
scaled photon energy $\om/Z_{\eff}^2$:
%%%%%%%% 5.29 %%%%%%%%%%%%%%%%%%%%%%%%%%%%
\begin{equation}
\d\sigma_{\pol}(\om, v_1) 
=
{N_{nl}^2 \over Z_{\eff}^2} \,
\d \sigma_{\pol}^{\rm red}
\left({\om\over Z_{\eff}^2},{v_1\over Z_{\eff}}\right). 
\label{scaling}
\end{equation}

In the limit of high-energy photons, $\om \gg I_{1s}$, the generalized 
polarizability is expressed via the form-factor $F(q)$
(see (\ref{1.4})).
The parameter $Z_{\eff}$ can be chosen equal to $Z$. 
Then, the PBrS cross section depends on the single parameter,
$\omega / (v\, Z)$, and the scaling law (\ref{scaling}) 
reduces to that discussed in \cite{IshiiMorita1986}.

%%%%%%%%%%%%%%%%
% Bethe ridge

As mentioned in section \ref{MainFeatures} there is a
peculiar feature in the PBrS emitted in 
the collision of a fast heavy charged particle with a many-electron atom. 
This feature originates from the kinematics of the atomic electrons 
in the process and  
manifests itself as an additional maximum in the velocity 
dependence of the PBrS cross section. 
A similar peculiarity, known as the Bethe ridge, appears 
in the differential cross section of the impact ionization of 
an atom by a charged particle \cite{Landau3}, where there is a ridge in 
the dependence of the cross section on the transferred momentum 
$q$ and the momentum of the outgoing electron $p_e$ at $q \sim p_e$. 
This ridge in the differential cross section is a result of the momentum 
transfer to one of the target electrons in the collision process. 
The peculiarity in the polarizational BrS process arises from the 
similar dynamics of atomic electrons 
\cite{KorolLyalinObolenskySolovyov2000}. 
However, in this process after virtual excitation the atomic electron 
returns to its initial state radiating a photon. 
The Bethe-type virtual excitations of electrons in the
PBrS process give rise to the additional maximum in the velocity 
dependence of the cross section.

%%%%%%%%%%%%%%%%%
\begin{figure}%[h]
\begin{center}
\includegraphics[width=10cm,height=10cm,angle=0]{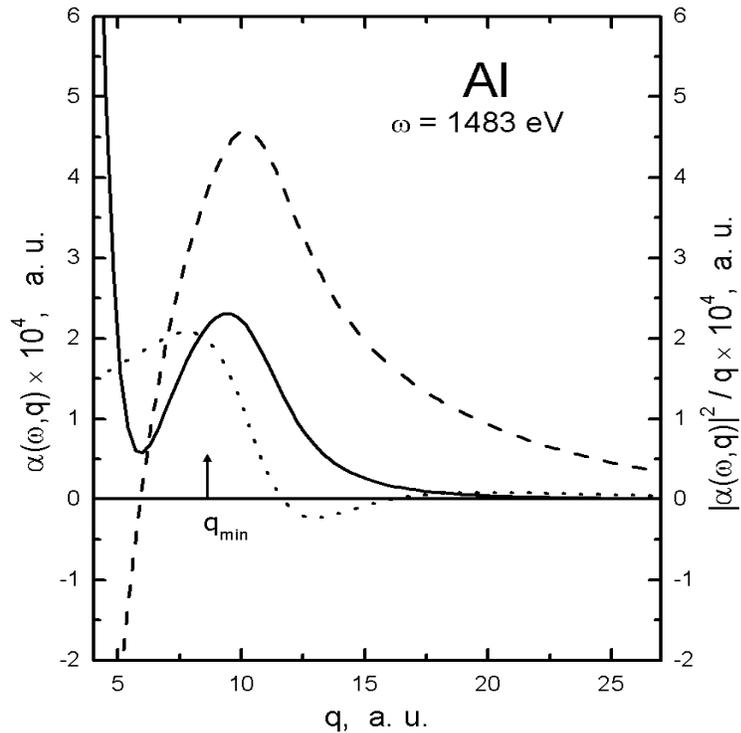}%{ridge1.eps}
\end{center}
\caption{The $q$-dependence of $\alpha(\omega,q)$ of an Al atom
at fixed frequency $\om=1483$ eV 
\protect\cite{KorolLyalinObolenskySolovyov2000}. 
Dashed and dotted curves represent
${\rm Re}\Bigl[\alpha(\omega,q)\Bigr]$ 
and
${\rm Im}\Bigl[\alpha(\omega,q)\Bigr]$ respectively.
The solid curve shows the behaviour of the integrand in
(\protect\ref{scaling.1}),
$\left| \alpha (\omega, q) \right|^2 /q$.}
\label{ridge.fig1}
\end{figure}
%%%%%%%%%%%%%%%%%
Let us briefly outline the reasons which lead to the 
 Bethe peculiarity in the velocity dependence of $\d\sigma_{\pol}(\om,v_1)$
(the details are given in \cite{KorolLyalinObolenskySolovyov2000}).
The cross section $\d\sigma_{\pol}(\om,v_1)$, defined in 
(\ref{scaling.1}), depends on the generalized polarizability
$\alpha(\om,q)$, which (in the case of a spherically-symmetric target)
can be written as follows:
\begin{equation}
\alpha(\om,q)
=
{1 \over (\bfe\bfq)}
\sum_n 
\left\{
{
\langle 0\left | \bfe\hat{\bfp} \right|n\rangle \, F_{n0}({\bfq})
\over \om_{n0} - \om - \i\,0
}
+
{
F_{0n}(\bfq)\, \langle n\left| \bfe\hat{\bfp} \right|0 \rangle
\over \om_{n0} + \om 
}
\right\}.
\label{alfadef}
\end{equation}
The notations are explained in the paragraph after eq. (\ref{1.8}).

The matrix element 
$F_{n0}(\bfq) = \langle m\left|\exp({\rm i}\bfq\bfr) \right|0\rangle$
in (\ref{alfadef})  as a function of the intermediate state energy $\E_n$
is maximal if $\E_n = q^2/2$. 
This means that the momentum is mainly transferred to the 
electron in the intermediate state.
Similar behaviour of matrix elements is known from electron--atom
impact ionization where it is called `the Bethe ridge'.
In the PBrS process the Bethe peculiarity results in the relationship 
between the photon energy and the transferred 
momentum:
\begin{equation} 
\om \approx {q^2 \over 2}+I, 
\label{ridgeomq}
\end{equation} 
where $I$ is the ionization potential of the subshell from 
which the electron is excited.
This relation defines the curve in the plane 
$(\omega,q)$ in the vicinity of which the generalized polarizability 
$\alpha(\omega,q)$ as a function of $q$ has a 
resonance character. This is illustrated by figure \ref{ridge.fig1}.

%%%%%%%%%%%%%%%%%
\begin{figure}%[h]
\begin{center}
\includegraphics[width=15cm,height=11cm,angle=0]{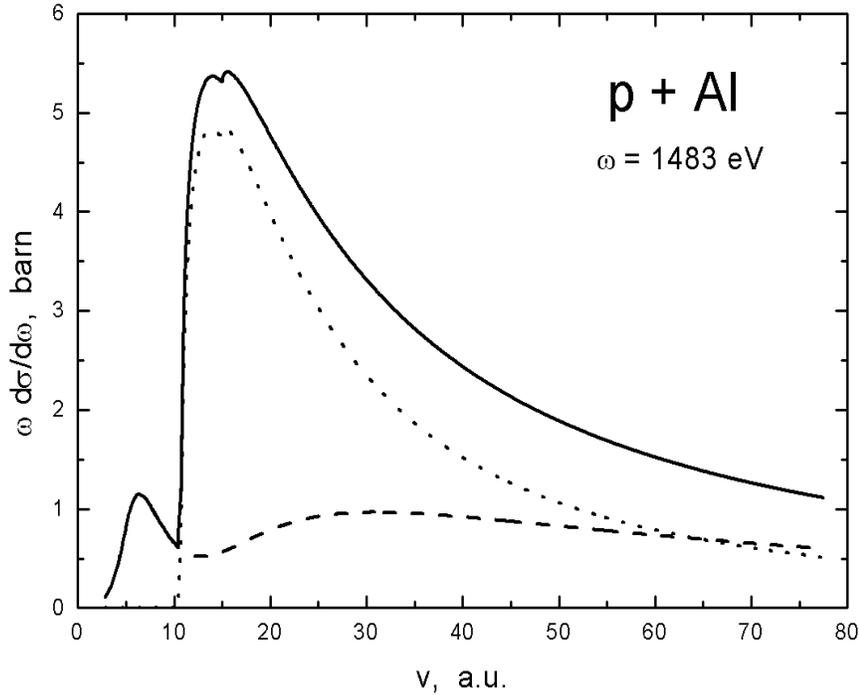}%{ridge2.eps}
\end{center}
\caption{
Velocity dependencies of the elastic PBrS (dotted curve) 
and and inelastic PBrS (dashed curve) cross sections 
at $\om = 1483$ eV.
The solid curve is the total photon emission cross section as a 
function of $v_1$ \protect\cite{KorolLyalinObolenskySolovyov2000}. }
\label{ridge.fig2}
\end{figure}
%%%%%%%%%%%%%%%%%

In \cite{KorolLyalinObolenskySolovyov2000} it was 
demonstrated that the resonant-like behavior 
of $\Bigl|\alpha(\omega,q)\Bigr|^2$
results in the peculiarity (a maximum) in the PBrS cross section  
$\d\sigma_{\pol}(\om,v_1)$ as a function of $v_1$.
The Bethe peculiarity is mostly pronounced 
if the velocity and the photon energy satisfy the conditions
$v_1 \sim \sqrt{(\omega + I)/2} < \sqrt{2\omega}$ which ensures
that the elastic PBrS dominates over the inelastic one in the total
emission spectrum, and, consequently, that the contribution of
inelastic channels does not smear out the peculiarity.

Figure \ref{ridge.fig2} illustrates this effect.
The maximum in the total emission spectrum (the solid curve) 
formed in $p+$Al collision is due to the maximum of 
$\d\sigma_{\pol}(\om,v_1)$ as a function of the velocity.
In the vicinity of maximum the contribution of the inelastic channel 
(the dashed curve) is small.

%%%%%%%%%%%%%%%%%%%%%%%%%%%%%%%%%%%%%%%%%%%%%%%%%%%%%%
\section{Conclusions}

This review was focused on the achievements made during last decade
in the development of the theory of polarizational BrS formed in 
non-relativistic collisions of various particles with {\em isolated} 
many-electron atoms. 
Apart from what has been described above in the paper we would like to 
mention that a number of algorithms and computer codes have been developed,
which allows one to perform an accurate quantitative analysis of the
spectral and spectral-angular distributions of the total BrS formed in 
an arbitrary collision process which involves a charged structureless
particle and a multi-electron complex (including molecules and clusters
\cite{RPC_Clusters2004}).

However, not all of the problems related to the PBrS effect have been
equally developed.
Among these we mention the process of inelastic BrS, the non-dipolar
effects, the role of PBrS in low-energy electron/positron--atom scattering.

The main physical reason for paying less attention to the process of 
inelastic BrS is that, as it was mentioned, in wide ranges of photon
energies and collision velocities the contribution of the inelastic
channels is parametrically smaller then that of the elastic BrS due
to the coherence nature of the latter.
However, the level of accuracy which can be achieved, at present, in
the quantitative description of the inelastic BrS, especially in the case
of electron-atom scattering, cannot be matched with that achieved
for the elastic BrS process.
An accurate study of the inelastic radiative scattering is a more 
difficult problem both from theoretical and computational points of view.  
Therefore there exist fewer in this field. 
Particular interest can be attached to the understanding of collective 
excitations in inelastic radiative scattering.

The discussions and numerical calculations of the
PBrS beyond the dipole-photon approximation have not been as extensive as
those based on the dipole-photon scheme.  
The non-dipolar effects include two aspects.
The first one is related to the emission into higher multipoles.
Its relative significance, which is governed by the magnitude
of the factor $k R_{\rm at}$ ($k=\om/c$ is the momentum of the photon),
increases with the photon energy $\om$. 
Therefore, if one is interested in the accurate data on the PBrS
characteristics in the range of photon energies higher than several keV,
the corrections to the dipole-photon approximation must be accounted
for.
The non-dipolar effects manifests themselves in PBrS also via the 
retardation in the interaction between the projectile and the target's
electrons. 
The retardation modifies the Coulomb interaction adding the terms
proportional to $v_1/c$ ($v_1$ is the velocity of the projectile).
These terms become noticeable for electrons of kinetic energy 
of several tens of keV.
We note that within this energy range the latest experiments on the
detection of PBrS were performed \cite{PortilloQuarles_PRL}.
Both of these effects, the emission into higher multipoles and the 
retardation, are incorporated in the full relativistic theory of
PBrS developed recently \cite{OurRelativisticJPB,OurRelativisticJETP}.

Finally, we mention the problem of the role of PBrS mechanism 
in a low-energy 
 ($\E_1<I_1$) electron--atom collision.
In this case the polarization of the target can influence the radiation
process not only via the PBrS mechanism directly but in a less evident
way.
Namely, the dynamically induced dipole moment may lead to the modification
of the projectile's wavefunction (more precisely, the phaseshifts).
This will influence the amplitude of the {\em ordinary} BrS.
This effect, which will be mostly pronounced for the targets with
large dipole polarizabilities, has not been studied in detail so far.

The above described processes do not cover all the phenomena which 
are worth to be investigated further.
The polarizational BrS problem is rather  broad,  
because this kind of radiation can be emitted in any collision involving
structured particles:  nuclei, atoms, molecules or clusters.
The number of various colliding pairs, different interaction
forces between particles, kinematical
conditions and the frequency ranges
make this problem quite  varied and interesting.  

%%%%%%%%%%%%%%%%%%%%%%%%%%%  Acknowledgment %%%%%%%%%%
\ack

This work is supported  by the Russian Foundation for Basic Research
(Grant No 96-02-17922-a) and INTAS (Grant No 03-51-6170).
AVK acknowledges the support of the Alexander von Humboldt Foundation.

%%%%%%%%%%%%%% references for the nrel.tex 

%%%%%%%%%%%%%%%%%%%%%%%%%%%%%%%%%%%%%%%%%%%%
%%%%%%%%%%%%%%%%%%%%%%%%%%%%%%%%%%%%%%%%%%%%
\end{document}